# Real-time fMRI neurofeedback training of the amygdala activity with simultaneous EEG in veterans with combat-related PTSD

Vadim Zotev[a], Raquel Phillips[a], Masaya Misaki[a], Chung Ki Wong[a], Brent E. Wurfel[a,b], Frank Krueger[c], Matthew Feldner[d], and Jerzy Bodurka[a,e,*]

[a]Laureate Institute for Brain Research, Tulsa, OK, USA; [b]Laureate Psychiatric Clinic and Hospital, Tulsa, OK, USA; [c]School of Systems Biology, George Mason University, Fairfax, VA, USA; [d]Dept. of Psychological Science, University of Arkansas, Fayetteville, AR, USA; [e]Stephenson School of Biomedical Engineering, University of Oklahoma, Tulsa, OK, USA

**Abstract:** Posttraumatic stress disorder (PTSD) is a chronic and disabling neuropsychiatric disorder characterized by insufficient top-down modulation of the amygdala activity by the prefrontal cortex. Real-time fMRI neurofeedback (rtfMRI-nf) is an emerging method with potential for modifying the amygdala-prefrontal interactions. We report the first controlled emotion self-regulation study in veterans with combat-related PTSD utilizing rtfMRI-nf of the amygdala activity. PTSD patients in the experimental group (EG, $n$=20) learned to upregulate blood-oxygenation-level-dependent (BOLD) activity of the left amygdala (LA) using the rtfMRI-nf during a happy emotion induction task. PTSD patients in the control group (CG, $n$=11) were provided with a sham rtfMRI-nf. The study included three rtfMRI-nf training sessions, and EEG recordings were performed simultaneously with fMRI. PTSD severity was assessed before and after the training using the Clinician-Administered PTSD Scale (CAPS). The EG participants who completed the study showed a significant reduction in total CAPS ratings, including significant reductions in avoidance and hyperarousal symptoms. They also exhibited a significant reduction in comorbid depression severity. Overall, 80% of the EG participants demonstrated clinically meaningful reductions in CAPS ratings, compared to 38% in the CG. No significant difference in the CAPS rating changes was observed between the groups. During the first rtfMRI-nf session, functional connectivity of the LA with the orbitofrontal cortex (OFC) and the dorsolateral prefrontal cortex (DLPFC) was progressively enhanced, and this enhancement significantly and positively correlated with the initial CAPS ratings. Left-lateralized enhancement in upper alpha EEG coherence also exhibited a significant positive correlation with the initial CAPS. Reduction in PTSD severity between the first and last rtfMRI-nf sessions significantly correlated with enhancement in functional connectivity between the LA and the left DLPFC. Our results demonstrate that the rtfMRI-nf of the amygdala activity has the potential to correct the amygdala-prefrontal functional connectivity deficiencies specific to PTSD.

**Keywords:** PTSD; Combat trauma; Neurofeedback; Real-time fMRI; EEG-fMRI; Amygdala; Dorsolateral prefrontal cortex; Orbitofrontal cortex; Functional connectivity; EEG coherence

## 1. Introduction

Posttraumatic stress disorder (PTSD) is a chronic and disabling neuropsychiatric disorder with lasting negative effects on personal well-being and high economic costs to the society (Kessler, 2000). Among veterans with PTSD who receive trauma-focused treatment, such as cognitive processing therapy or prolonged exposure therapy, only 50% to 70% achieve clinically meaningful symptom improvement, and as many as 66% retain their PTSD diagnosis after treatment (Steenkamp et al., 2015).

Treatment of PTSD is complicated by the fact that this disorder afflicts functions of several brain systems (Liberzon & Abelson, 2016). First, abnormalities in the salience network (SN) function are associated with exaggerated threat detection. This network includes the amygdala, the insula, the dorsal anterior cingulate cortex (dACC), and other regions. Second, abnormal functioning of the executive function/emotion regulation (EF/ER) system leads to cognitive and emotion regulation impairments. This system includes regions of the prefrontal cortex (PFC): the dorsolateral PFC (DLPFC), the ventrolateral PFC (VLPFC), and the medial PFC (MPFC). The MPFC, in turn, includes the ventromedial PFC (VMPFC), the dorsomedial PFC (DMPFC), the rostral ACC (rACC), and other regions. Third, functional deficiencies in the brain circuits involved in contextual processing (CP) lead to difficulties in threat discrimination. These circuits include the hippocampus, the thalamus, the locus coeruleus, and the MPFC areas (Liberzon & Abelson, 2016).

Functional neuroimaging studies of emotional processing have demonstrated prominent involvement of the above-mentioned brain systems (SN, EF/ER, CP) in neurobiology of PTSD (e.g. Etkin & Wager, 2007; Lanius et al., 2006; Patel et al., 2012; Shin & Liberzon, 2010). In particular, numerous studies have shown hyperactivity of the amygdala and hypoactivity of the PFC regions during emotional processing in PTSD patients compared to control participants (e.g. Etkin & Wager, 2007). This finding is often interpreted as indicating an insufficient top-down regulation of the amygdala activity by the PFC. It also suggests that functional connectivity between the amygdala and the PFC is abnormally reduced in PTSD. Among the PFC regions, the orbitofrontal cortex (OFC), along with the rACC and the subgenual ACC (sgACC), has the densest neuronal connections to the amygdala (Ghashghaei et al., 2007). The lateral OFC (LOFC, BA 47, part of the VLPFC, as well as lateral BA 11) and the medial OFC (MOFC, BA 11, part of the MPFC adjacent to the VMPFC) play important roles in emotion regulation and reward/punishment-guided learning (Kringelbach & Rolls, 2004; Ochsner et al., 2002; Rushworth et al., 2011). Abnormalities in the LOFC and MOFC functions are observed in anxiety disorders (Milad & Rauch, 2007), including hypoactivity of these regions in PTSD (Lanius et al., 2006; Patel et al., 2012).

*Corresponding author. E-mail: jbodurka@laureateinstitute.org

Real-time fMRI neurofeedback (rtfMRI-nf) is a promising neuromodulation technique that allows non-invasive volitional modulation of blood-oxygenation-level-dependent (BOLD) activity of small precisely defined regions deep inside the brain (e.g. Birbaumer et al., 2013; Sulzer et al., 2013; Thibault et al., 2016; Weiskopf, 2012). We demonstrated previously that the rtfMRI-nf training of the amygdala activity enhanced both functional and effective connectivities between the amygdala and the PFC (Zotev et al., 2011, 2013). This is not surprising, because an rtfMRI-nf training in general is a goal-oriented behavior that engages the EF/ER system (Zotev et al., 2016). With the amygdala as the target region, the rtfMRI-nf training has the potential to enhance top-down modulation of the amygdala activity by the EF/ER. The rtfMRI-nf of the amygdala activity has been shown to reduce depressive symptoms in patients with major depressive disorder (MDD) (Young et al., 2017; Zotev et al., 2016). In addition, EEG recordings during the rtfMRI-nf procedure revealed left-lateralized enhancement in EEG coherence that positively correlated with depression severity (Zotev et al., 2016). This finding suggests that the rtfMRI-nf of the amygdala activity can correct (reverse) the functional connectivity abnormalities specific to MDD, and, possibly, other neuropsychiatric disorders.

Two recent pilot studies explored the feasibility of using rtfMRI-nf of the amygdala activity for treatment of PTSD. The first pilot study (Nicholson et al., 2017) involved ten PTSD patients. The participants learned to downregulate BOLD activity of the bilateral amygdala using rtfMRI-nf while viewing personalized trauma words. The study included one rtfMRI-nf session, and changes in PTSD severity were not assessed. Increased activations in the DLPFC and LOFC, as well as enhanced fMRI connectivity of the amygdala with the DLPFC and DMPFC, were observed during the rtfMRI-nf task compared to a control task (Nicholson et al., 2017). The second pilot study (Gerin et al., 2016) included three combat veterans with chronic PTSD. The participants used rtfMRI-nf to downregulate BOLD activity of the bilateral amygdala after listening to a personal trauma-based audio script. After three rtfMRI-nf sessions, two participants showed clinically meaningful reductions in PTSD severity. Increased resting fMRI connectivity of the amygdala with the MOFC and rACC/sgACC was observed after the training (Gerin et al., 2016). Clearly, both studies had limited statistical powers due to the small sample sizes. Moreover, neither study included a control group, so specificity of the reported effects to the amygdala-based rtfMRI-nf could not be verified.

Here we report results from the first controlled rtfMRI-nf study of emotion self-regulation in veterans with combat-related PTSD. In our study, PTSD patients learned to upregulate BOLD activity of the left amygdala (LA) using rtfMRI-nf while performing a positive emotion induction task we introduced earlier (Zotev et al., 2011). EEG recordings were conducted simultaneously with the rtfMRI-nf procedure to explore its electrophysiological correlates (Zotev et al., 2016, 2018). We tested the following hypotheses regarding effects of the rtfMRI-nf training targeting the LA activity. First, we hypothesized that neurofeedback-naïve PTSD patients would be able to significantly increase BOLD activity of the LA during the training. Second, we hypothesized that most participants would achieve clinically meaningful reductions in PTSD severity (as explained below) at the end of the study. Third, we hypothesized that fMRI functional connectivity of the LA with key PFC regions involved in emotion regulation would be enhanced during the training and these enhancements would correlate with PTSD severity. We also expected to observe differences between effects of the LA-based rtfMRI-nf and those of sham rtfMRI-nf. We could not hypothesize on significance of such differences due to the lack of prior controlled studies using rtfMRI-nf in PTSD.

## 2. Methods

### 2.1. Study overview

The study was conducted at the Laureate Institute for Brain Research, and was approved by the Western Institutional Review Board (IRB). All study procedures were carried out in accordance with the principles expressed in the Declaration of Helsinki.

The study included eight sessions (visits), illustrated schematically in Fig. 1A. The visits typically followed with one week intervals. Each visit involved a psychological evaluation by a licensed psychiatrist in addition to experimental procedures.

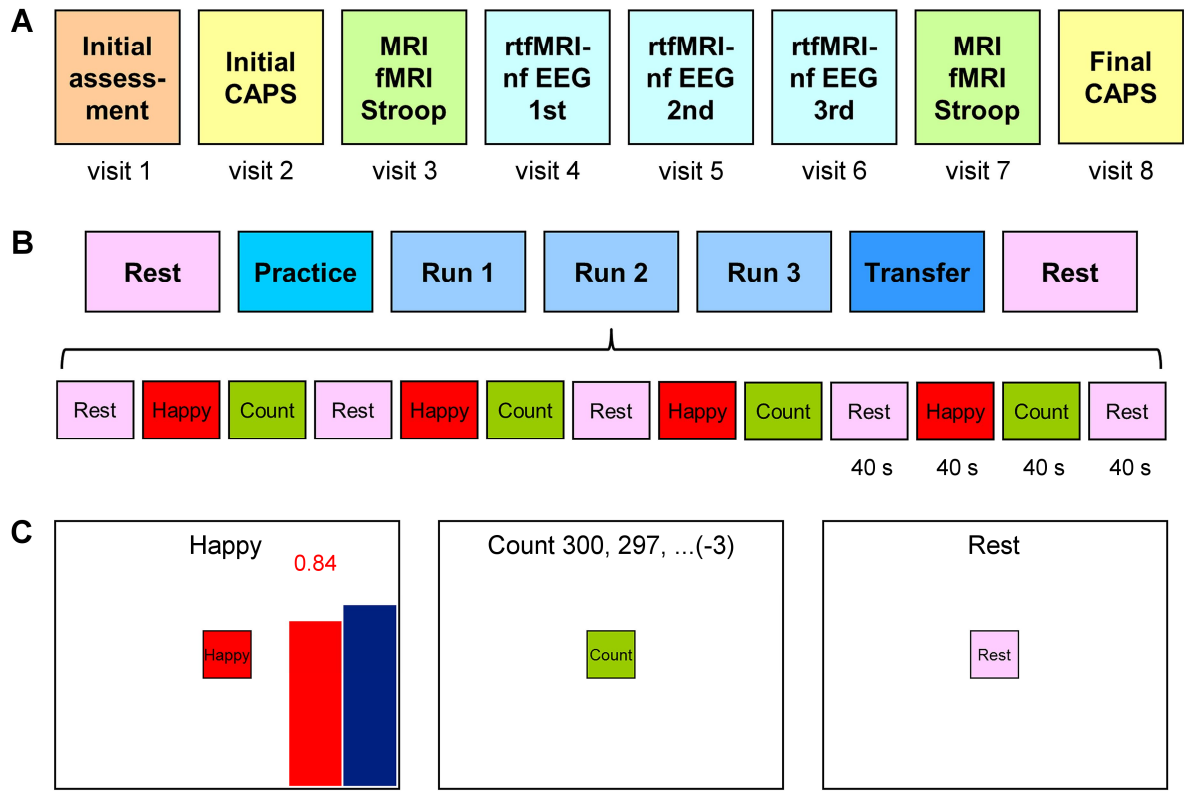

**Fig. 1.** Overview of the emotion self-regulation study utilizing real-time fMRI neurofeedback (rtfMRI-nf) of the amygdala in veterans with combat-related PTSD. **A)** The study included eight sessions (visits) with three rtfMRI-nf training sessions (visits 4, 5, 6) and two PTSD symptom assessment (CAPS) sessions (visits 2, 8). **B)** Experimental protocol for one rtfMRI-nf session. It consisted of seven runs, each lasting 8 min 46 s. It included two Rest runs, four rtfMRI-nf runs – Practice, Run 1, Run 2, Run 3, and a Transfer run without nf. The names of the five task runs are abbreviated in the text and figures as PR, R1, R2, R3, and TR, respectively. The experimental runs (except the Rest) consisted of 40-s long blocks of Happy Memories, Count, and Rest conditions (abbreviated as H, C, and R, respectively). **C)** Real-time GUI display screens for the Happy Memories, Count, and Rest conditions. The rtfMRI-nf signal is displayed during the Happy Memories conditions in the four nf runs as the variable-height red bar. The height of the red bar represents real-time fMRI activity of the target ROI. It is updated every 2 s. The height of the blue bar specifies a target level for the rtfMRI-nf signal. It is raised from run to run.

Visit 1 was the initial assessment visit. It included administration of the following tests: the Edinburgh Handedness Inventory (EHI, Oldfield, 1971), the Family Interview for Genetic Studies (FIGS, Maxwell, 1992), the Fagerström Test for Nicotine Dependence (FTND, Fagerström, 1978), the Hollingshead Four-factor Index of Socioeconomic Status (SES, Hollingshead, 1975), the Quick Inventory of Depressive Symptomatology (QIDS, Rush et al., 2000), and the 21-item Hamilton Depression Rating Scale (HDRS, Hamilton, 1960).

Visit 2 included the initial assessment of PTSD severity by means of the Clinician-Administered PTSD Scale for DSM-IV (CAPS, Blake et al., 1990; Weathers et al., 2001). It also included completion of the 20-item Toronto Alexithymia Scale (TAS-20, Bagby et al., 1994), and the Emotion Contagion Scale (EC, Doherty, 1997).

Visit 3 included the emotional counting Stroop task (ecStroop, Whalen et al., 2006) with simultaneous fMRI, and the Script-Driven Imagery Procedure (SDIP, Pitman et al., 1987) with the Responses to Script-Driven Imagery Scale (RSDI, Hopper et al., 2007). At the beginning of the visit, the HDRS, the Montgomery-Asberg Depression Rating Scale (MADRS, Montgomery & Asberg, 1979), the PTSD CheckList Military Version (PCL-M, Weathers et al., 1991), the Hamilton Anxiety Rating Scale (HARS, Hamilton, 1959), and the Snaith-Hamilton Pleasure Scale (SHAPS, Snaith et al., 1995) were administered. The Profile of Mood States (POMS, McNair et al., 1971) and the Visual Analog Scale (VAS, with 10-point subscales for happy, restless, sad, anxious, irritated, drowsy, and alert states) were completed by participants both before and after the ecStroop and the SDIP procedures.

Visits 4, 5, and 6 were the neurofeedback training sessions each involving the rtfMRI-nf with simultaneous EEG procedure, illustrated in Fig. 1B. At the beginning of each session, the HDRS, the MADRS, the HARS, the PCL-M, and the SHAPS scales were administered. The POMS and the VAS were completed both before and after the rtfMRI-nf procedure in each visit.

Visit 7 included the same procedures as visit 3.

Visit 8 included the final assessments of PTSD severity using the CAPS.

### 2.2. Participants

All the participants provided a written informed consent as approved by the IRB. They met the criteria for PTSD specified in the *Diagnostic and Statistical Manual of Mental Disorders, Fourth Edition Text Revision* (DSM-IV-TR) (American Psychiatric Association, 2000). All the participants were male and had PTSD related to combat trauma as their primary diagnosis. They received monetary compensation for their participation in the study.

The participants were randomly assigned to either the experimental group (EG) or the control group (CG) at approximately 2:1 ratio. They were unaware of their group status. During the rtfMRI-nf training sessions (visits 4, 5, 6), the EG participants were provided with rtfMRI-nf based on BOLD activity of the LA (Zotev et al., 2011). The CG participants were provided, without their knowledge, with sham rtfMRI-nf based on BOLD activity of a control region, presumably not involved in emotion processing (Zotev et al., 2011). Selection of the target regions for rtfMRI-nf is described in detail below. Participants in both groups received identical instructions and followed the same procedures.

Table 1 reports main characteristics of the EG and CG groups. In the EG, 20 participants completed the first rtfMRI-nf session (visits 1-4), and 15 of them completed the whole study (visits 1-8, Fig. 1A). In the CG, 11 subjects completed the first rtfMRI-nf session, and eight of them completed the whole study. Mean PTSD severity ratings (CAPS, PCL-M) and comorbid depression severity ratings (HDRS, MADRS) are specified for each group in Table 1. There were no significant group differences in age, CAPS, PCL-M, HDRS, or MADRS between the EG and CG at the beginning of the study.

### 2.3. Experimental protocol

The experimental protocol for each rtfMRI-nf session (Fig. 1B) was similar to the one we employed previously in rtfMRI-nf studies with healthy participants (Zotev et al., 2011, 2014) and MDD patients (Young et al., 2017; Zotev et al., 2016). Prior to each rtfMRI-nf session, a participant was given detailed instructions that included an overview of the experiment and an explanation of each experimental task. The participant was asked to think of and write down five happy autobiographical memories. It was suggested that he use those memories at the beginning of the rtfMRI-nf training to evaluate their effects, and then explore various other happy autobiographical memories as the training progressed to enhance happy emotion and improve rtfMRI-nf performance.

Each rtfMRI-nf session included seven fMRI runs (Fig. 1B), and each run lasted 8 min 46 s. During the initial and final Rest runs, the participants were asked to relax and rest while looking at a fixation cross. The five task runs – the Practice run, Run 1, Run 2, Run 3, and the Transfer run – consisted of alternating 40-s long blocks of Happy Memories, Count, and Rest conditions (Fig. 1B). The real-time GUI display screens for these conditions are shown schematically in Fig. 1C. Each condition was specified by visual cues that included a colored square with the condition name at the center of the screen and a text line at the top of the screen. For the Happy Memories condition blocks, the participants were instructed to feel happy by evoking and contemplating happy autobiographical memories while simultaneously trying to raise the variable-height red rtfMRI-nf bar on the screen to the target level of the blue bar (Fig. 1C, left). The red bar height was updated every 2 s, and was also indicated by the red numeric value shown above the bar (Fig. 1C, left). For the Count condition blocks, the subjects were instructed to mentally count back from 300 by subtracting a given integer as shown on the screen (Fig. 1C, middle). For the Rest condition blocks, the participants were asked to rest and let their minds wander while looking at the screen (Fig. 1C, right).

During the four rtfMRI-nf runs (Practice, Runs 1-3), the participants performed the three experimental tasks as indicated by the GUI display screens shown in Fig. 1C. The target level for the rtfMRI-nf (blue bar in Fig. 1C, left) was fixed during each run, but was raised in a linear fashion across the four nf runs. It was set to 0.5%, 1.0%, 1.5%, and 2.0% for the Practice run, Run 1, Run 2, and Run 3, respectively (see Fig. 3A below). During the Practice run, the participants were given an opportunity to become familiar with (or refresh knowledge of) the rtfMRI-nf procedure and to evaluate emotional impact of the happy autobiographical memories they had prepared. During the Transfer run, the participants performed the same tasks as during the preceding nf runs, except that no bars were shown on the screen during the Happy Memories conditions, and the text line read "As Happy as possible". The Transfer run was included to evaluate whether the participants' learned ability to control BOLD activity of the target ROI generalized beyond the actual rtfMRI-nf training when the nf information was no longer provided. The Count conditions involved counting back from 300 by subtracting 3, 4, 6, 7, and 9 for the Practice run, Run 1, Run 2, Run 3, and the Transfer run, respectively. After each experimental run with the Happy Memories task, a participant was asked to verbally rate his performance on a scale from 0 ("not at all") to 10 ("extremely") by answering two questions: "How successful were you at recalling your happy memories?" and "How happy are you right now?".

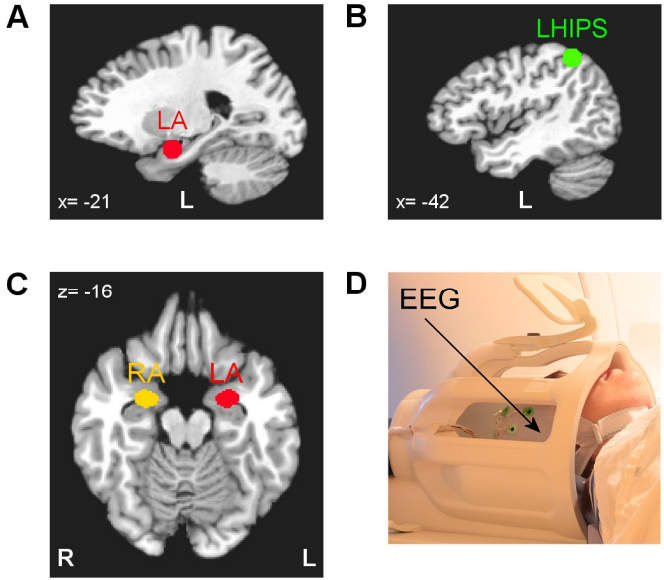

**Fig. 2.** Regions of interest for real-time fMRI data processing and offline fMRI data analyses. **A)** Spherical 14-mm diameter target ROI in the left amygdala (LA) region used to provide rtfMRI-nf for the experimental group (EG). **B)** Spherical 14-mm diameter target ROI in the left horizontal segment of the intraparietal sulcus (LHIPS) region used to provide sham rtfMRI-nf for the control group (CG). **C)** Left amygdala (LA) and right amygdala (RA) ROIs defined anatomically as the amygdala regions specified in the co-planar stereotaxic atlas of the human brain by Talairach and Tournoux. These ROIs were employed in the offline fMRI data analyses. The ROIs are projected in the figure onto the standard TT_N27 template in the Talairach space. Following the radiological notation, the left hemisphere (L) is shown to the reader's right. **D)** A 32-channel MR-compatible EEG system from Brain Products, GmbH was used to perform EEG recordings during fMRI.

### 2.4. Regions of interest

The rtfMRI-nf procedure was based on the target region-of-interest (ROI) approach we employed previously (Zotev et al., 2011, 2016). Two target ROIs were defined as 14-mm diameter spheres in the stereotaxic array of Talairach and Tournoux (Talairach & Tournoux, 1988). The target ROI centered at (−21, −5, −16) in the left amygdala (LA) region (Fig. 2A) was used for the EG. The target ROI centered at (−42, −48, 48) in the left horizontal segment of the intraparietal sulcus (LHIPS) region (Fig. 2B) was used for the CG. The specified ROI centers were selected based on quantitative meta-analyses of functional neuroimaging studies investigating the role of the amygdala in emotion processing (Sergerie et al., 2008) or the role of the HIPS in number processing (Dehaene et al., 2003). During the experiment, these target ROIs were transformed from the Talairach space to each participant's individual fMRI (EPI) image space and used to provide rtfMRI-nf signal depending on the group. For offline fMRI data analysis, the left amygdala (LA) and the right amygdala (RA) ROIs (Fig. 2C) were defined anatomically as the amygdala regions specified in the Talairach-Tournoux brain atlas in AFNI (Cox, 1996; Cox & Hyde, 1997).

### 2.5. Data acquisition

All experiments were conducted on the General Electric Discovery MR750 3T MRI scanner with a standard 8-channel receive-only head coil (Fig. 2D). A single-shot gradient echo EPI sequence with FOV/slice=240/2.9 mm, $TR/TE$=2000/30 ms, flip angle=90°, 34 axial slices per volume, slice gap=0.5 mm, SENSE $R$=2 in the phase encoding (anterior-posterior) direction, acquisition matrix 96×96, sampling bandwidth=250 kHz, was employed for fMRI. Each fMRI run lasted 8 min 46 s and included 263 EPI volumes (the first three EPI volumes were included to allow fMRI signal to reach a steady state and were excluded from data analysis). Physiological pulse oximetry and respiration waveforms were recorded simultaneously with fMRI. The EPI images were reconstructed into a 128×128 matrix, resulting in 1.875×1.875×2.9 mm$^3$ fMRI voxels. A T1-weighted 3D MPRAGE sequence with FOV/slice=240/1.2 mm, $TR/TE$=5.0/1.9 ms, $TD/TI$= 1400/725 ms, flip angle=10°, 128 axial slices per slab, SENSE $R$=2, acquisition matrix 256×256, sampling bandwidth=31.2 kHz, scan time=4 min 58 s, was used for structural imaging. It provided high-resolution anatomical brain images with 0.94×0.94×1.2 mm$^3$ voxels.

EEG recordings were performed simultaneously with fMRI (Fig. 2D) using a 32-channel MR-compatible EEG system from Brain Products, GmbH. The EEG system clock was synchronized with the MRI scanner 10 MHz clock using the Brain Products' SyncBox device. EEG data were acquired with 0.2 ms temporal and 0.1 μV measurement resolution (16-bit 5 kS/s sampling) in 0.016...250 Hz frequency band with respect to FCz reference. All technical details of the EEG-fMRI system configuration and data acquisition were reported previously (Zotev et al., 2012). Similar to our recent works (Zotev et al., 2016, 2018), the EEG recordings in the present study were passive, i.e. no EEG information was used in real time as part of the experimental procedure.

### 2.6. Real-time data processing

The rtfMRI-nf was implemented using the custom real-time fMRI system utilizing real-time functionality of AFNI (Cox, 1996; Cox & Hyde, 1997) as described previously (Zotev et al., 2011). A high-resolution MPRAGE anatomical brain image and a short EPI dataset (5 vol-umes) were acquired prior to each rtfMRI-nf session. The last volume in the EPI dataset was used as a reference EPI volume defining the subject's individual EPI space. The LA and LHIPS target ROIs, defined in the Talairach space (Fig. 2A,B) were transformed to the individual EPI space using the MPRAGE image data. The resulting ROIs in the EPI space contained approximately 140 voxels each. During the subsequent fMRI runs (Fig. 1B), the AFNI real-time plugin was used to perform volume registration of each acquired EPI volume to the reference EPI volume (motion correction) and export mean values of fMRI signals for these ROIs in real time. The custom developed GUI software was used to further process the exported fMRI signal values and display the ongoing rtfMRI-nf information (Fig. 1C). The rtfMRI signal for each Happy Memories condition was computed as a percent signal change relative to the baseline obtained by averaging fMRI signal values for the preceding Rest condition block (Fig. 1B). A moving average of the current and two preceding rtfMRI signal values was computed to reduce effects of fMRI noise and physiological artifacts (Zotev et al., 2011). This average value was used to set the height of the red rtfMRI-nf bar (Fig. 1C) every $TR$=2 s.

### 2.7. fMRI data analysis

Offline analysis of the fMRI data was performed in AFNI as described in detail in *Supplementary material* (S1.1). The analysis involved fMRI pre-processing with despiking, cardiorespiratory artifact correction (Glover et al., 2000), slice timing correction, and volume registration. A general linear model (GLM) fMRI activation analysis with Happy Memories and Count block-stimulus conditions was applied to the preprocessed fMRI data. Average GLM-based fMRI percent signal changes were computed for the anatomical LA ROI (Fig. 2C).

### 2.8. fMRI connectivity analyses

Analyses of fMRI functional connectivity for the LA as the seed region were performed within the GLM framework. The fMRI data were bandpass filtered between 0.01 Hz and 0.08 Hz. The six fMRI motion parameters were similarly filtered. The LA ROI (Fig. 2C) was transformed to each subject's individual high-resolution anatomical image space, and then to the individual EPI image space. The LA ROI in the EPI space included ~100 voxels. In addition, 10-mm-diameter ROIs were defined within the left and right frontal white matter (WM) and within the left and right ventricle cerebrospinal fluid (CSF). These ROIs were defined using individual high-resolution anatomical brain maps and similarly transformed. The resulting ROIs in the individual EPI space were used as masks to obtain average time courses for the LA, left and right WM, and left and right CSF regions. A single-subject GLM-based functional connectivity analysis was conducted for each task run using the 3dDeconvolve AFNI program. The -censor option was used to restrict the analysis to the Happy Memories condition blocks in each run. The GLM model included the time course of the LA ROI as the stimulus (seed) regressor. Nuisance covariates included five polynomial terms, time courses of the six fMRI motion parameters (together with the same time courses shifted by one $TR$), time courses of the left and right WM and CSF ROIs to reduce physiological noise (Jo et al., 2010), and step functions to account for the breaks in the data between the Happy Memories condition blocks. Each GLM analysis provided $R^2$-statistics and $t$-statistics maps for the stimulus regressor term, which we used to compute the correlation coefficient for each voxel. The correlation coefficient maps were Fisher $r$-to-$z$ normalized, transformed to the Talairach space, and re-sampled to $2\times2\times2$ mm$^3$ isotropic voxel size. The resulting individual LA fMRI connectivity maps were spatially smoothed (5 mm FWHM) and submitted to group analyses.

Three different group analyses were conducted for the LA fMRI connectivity data and the corresponding psychological data, separately for the EG and CG.

First, an analysis of correlations between the LA fMRI connectivity during the Happy Memories conditions with rtfMRI-nf and initial PTSD severity was performed for the Practice run in the 1st rtfMRI-nf session (visit 4, Fig. 1A). Group analysis on the LA fMRI connectivity data was conducted using the 3dttest++ AFNI program. It included three covariates: the initial CAPS ratings (visit 2), the corresponding HDRS ratings (visit 3), and the average individual fMRI connectivity of the LA with central WM. The last covariate accounted for residual spurious LA connectivity effects caused e.g. by head motion. The central WM mask was defined using the standard AFNI white matter mask in the Talairach space (TT_wm), which was re-sampled to $2\times2\times2$ mm$^3$ voxels, subjected to three-step erosion, and limited to 15<$z$<35 mm slab. The individual-subject LA connectivity values were averaged within this WM mask to yield a single covariate value for each subject. Centering of the three covariates was performed within the 3dttest++ program by subtraction of their mean values. The LA fMRI connectivity vs CAPS correlation effect was the main effect of interest.

Second, an analysis of correlations between the LA fMRI connectivity enhancement across the four neurofeedback runs in the 1st rtfMRI-nf session (visit 4, Fig. 1A) and initial PTSD severity was conducted as follows. An fMRI connectivity slope (FCS) was defined for each voxel as a slope of a linear trend in fMRI connectivity with the LA seed ROI across the Happy Memories conditions in the four nf runs (Practice, Run 1, Run 2, Run 3), as illustrated in Fig. 3B. The LA fMRI connectivity maps in the Talairach space for the four nf runs were concatenated, and the 3dTfitter AFNI program was used to carry out a voxel-wise linear trend analysis, yielding an FCS map for each subject. Group analysis on the LA FCS data was carried out using the 3dttest++ AFNI program. It included three covariates: the initial CAPS ratings (visit 2), the corresponding HDRS ratings (visit 3), and the average individual LA FCS for central WM. The last covariate was computed using the same central WM mask as described above, and accounted for spurious LA connectivity trends across the four nf runs. The FCS vs CAPS correlation effect and the mean FCS effect were the main effects of interest.

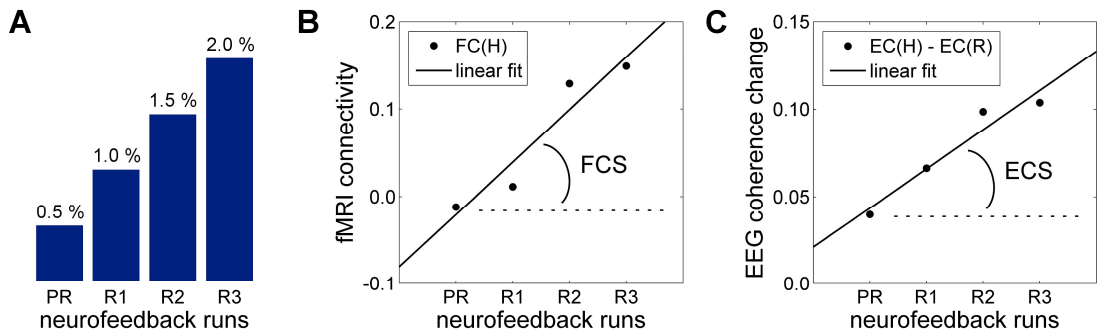

**Fig. 3.** Measures used to characterize linear trends in functional connectivity across neurofeedback runs in fMRI and EEG domains. **A)** The target level for the rtfMRI-nf (blue bar in Fig. 1C) was raised in a linear fashion across the four rtfMRI-nf runs (Practice, Run 1, Run 2, Run 3) in each neurofeedback session (Fig. 1B). **B)** Definition of the fMRI connectivity slope (FCS). It is defined, for each fMRI voxel, as a slope of a linear trend in fMRI connectivity during the Happy Memories conditions, FC(H), with the seed ROI across the four rtfMRI-nf runs. **C)** Definition of the EEG coherence slope (ECS). It is defined, for each pair of EEG channels, as a slope of a linear trend in upper alpha EEG coherence changes between the Rest and Happy Memories conditions, EC(H)−EC(R), across the four rtfMRI-nf runs.

Third, an analysis of correlations between the LA fMRI connectivity changes between the initial and final rtfMRI-nf sessions and the corresponding changes in PTSD severity was performed as follows. The LA fMRI connectivities during the Happy Memories conditions were averaged across the four nf runs (Practice, Run 1, Run 2, Run 3) in the initial (1st, visit 4) and in the final (3rd, visit 6) rtfMRI-nf sessions (Fig. 1A), and their voxel-wise differences (final vs initial) were computed. Changes between the final (visit 8) and initial (visit 2) CAPS ratings were considered for each participant, as well as changes between the final (visit 7) and initial (visit 3) HDRS ratings. Group analysis on the LA fMRI connectivity changes was conducted using the 3dttest++ AFNI program. It included three covariates: the changes in CAPS ratings, the changes in HDRS ratings, and the average individual changes in LA connectivity with central WM. The last covariate was determined using the same central WM mask as described above. The LA fMRI connectivity changes vs CAPS changes correlation effect and the mean fMRI connectivity changes between the sessions were the main effects of interest.

In each of these group analyses, statistical results were corrected for multiple comparisons by controlling the family-wise-error (FWE). The correction was based on Monte Carlo simulations implemented in the AlphaSim AFNI program.

*2.9. EEG data analysis*

Offline analysis of the EEG data, acquired simultaneously with fMRI, was performed using BrainVision Analyzer 2.1 software (Brain Products, GmbH) as described in detail in *Supplementary material* (S1.2). Removal of EEG artifacts was based on the average artifact subtraction and independent component analysis (Bell & Sejnowski, 1995; McMenamin et al., 2010). Channel Cz was selected as a new reference, and FCz was restored as a regular channel. Following the artifact removal, data from 29 EEG signal channels were downsampled to 8 ms temporal resolution. The upper alpha EEG band was defined individually for each participant as [IAF...IAF+2] Hz, where IAF is the individual alpha peak frequency. The IAF was determined by inspection of average EEG spectra for the occipital and parietal EEG channels across the Rest condition blocks in the four rtfMRI-nf runs (Fig. 1B).

*2.10. EEG coherence analysis*

EEG coherence analyses were conducted separately for the Rest and Happy Memories conditions in each of the four rtfMRI-nf runs (Fig. 1B). Each analysis included a segmentation with 4.096 s intervals (with exclusion of bad intervals, see S1.2), a complex FFT with 0.244 Hz spectral resolution, and the Coherence transform implemented in the Analyzer 2.1. A coherence value for signals from two EEG channels at a given frequency was computed as the squared magnitude of their cross spectrum value normalized by their power spectrum values at the same frequency ('magnitude-squared coherence' method). An average coherence value for the individual upper alpha EEG band [IAF...IAF+2] Hz was computed for each channel pair.

An analysis of correlations between the EEG coherence enhancement across the four neurofeedback runs in the 1st rtfMRI-nf session (visit 4, Fig. 1A) and initial PTSD severity was carried out as follows. An EEG coherence slope (ECS) was defined for each channel pair as a slope of a linear trend in the upper alpha EEG coherence changes between the Rest and Happy Memories conditions across the four nf runs (Practice, Run 1, Run 2, Run 3), as illustrated in Fig. 3C. (The changes between conditions were considered to reduce effects of any residual EEG artifacts). Analysis of partial correlations between the ECS values and the initial CAPS ratings (visit 2), controlled for the corresponding HDRS ratings (visit 3), was performed using the partialcorr() function in MATLAB.

## 2.11. Statistical tests

Inferential statistical analyses were performed in IBM SPSS Statistics 20 and MATLAB Statistics toolbox. To compare rtfMRI-nf effects between the EG and CG, we applied a two-way 4 (Training) × 2 (Group) between-within mixed factorial repeated measures ANOVA on fMRI percent signal changes (or, alternatively, fMRI connectivity measures) for a given ROI with Training (PR, R1, R2, R3) as a within-subject factor and Group (EG, CG) as a between-subject factor. Group statistics for individual runs were evaluated using $t$-tests. Effect sizes were characterized using Cohen's $d$. Correction for multiple comparisons was based on controlling the false discovery rate (FDR $q$), which was computed by applying the 3dFDR AFNI program to a column of uncorrected $p$-values from multiple tests.

## 3. Results

### 3.1. Psychological measures

Changes in PTSD severity and comorbid depression severity for the veterans who completed the study are reported in Table 2. The initial and final CAPS ratings were assessed during visits 2 and 8, respectively (Fig. 1A). The initial and final HDRS ratings were determined during visits 3 and 7, respectively. The participants in the EG showed a significant reduction in the total CAPS ratings after the study (EG: $t(14)=-3.69$, $p<0.0024$, $q<0.004$), with significant reductions in sub-scores for avoidance symptoms (EG: $t(14)=-3.78$, $p<0.0020$, $q<0.004$) and hyperarousal symptoms (EG: $t(14)=-2.54$, $p<0.024$, $q<0.030$). The EG participants also exhibited a significant reduction in the HDRS ratings (EG: $t(14)=-4.61$, $p<0.0004$, $q<0.002$). The participants in the CG similarly showed reductions in the total CAPS and HDRS ratings after the study, which, however, were not significant with smaller effect sizes (Table 2). Individual PTSD severity changes are illustrated in *Supplementary material* (S2.1, Fig. S1).

In the EG, 12 participants out of 15 (i.e. 80%) demonstrated clinically meaningful reductions in CAPS ratings (by 10 points or more) at the end of the study. In the CG, 3 subjects out of 8 (or 38%) showed clinically meaningful CAPS reductions. However, no significant difference in the CAPS rating changes (final vs initial) was observed between the two groups (EG vs CG: $t(21)=-0.90$, $p<0.377$, $d=-0.40$). Similarly, the HDRS rating changes (final vs initial) showed no significant group difference (EG vs CG: $t(21)=-0.22$, $p<0.825$, $d=-0.10$).

Comparison of the initial clinical ratings for the participants who completed the study and those who dropped out without completion (*Supplementary material* S2.1, Table S1) suggested that the subjects with higher initial PTSD severity were more likely to stay on and complete the study in either group (EG, CG).

### 3.2. Amygdala BOLD activity

Fig. 4 shows results of the offline fMRI activation analyses for the LA ROI (Fig. 2C) across the three rtfMRI-nf sessions (Fig. 1A). In the EG, the numbers of participants who completed the 1st, the 2nd, and the 3rd nf sessions were 20, 20, and 18, respectively. In the CG, the numbers of subjects who completed these sessions were 11, 10, and 10, respectively. One EG participant, who consistently showed negative LA fMRI activations during the three nf sessions, was considered an outlier, and this participant's results were excluded from the analyses reported in this section. Results from all the other EG and CG participants were included.

During the 1st rtfMRI-nf session, the LA BOLD activity for the Happy Memories conditions for the EG (H vs R, Fig. 4A, left) was significant after FDR correction ($q<0.05$) for Run 3 (R3: $t(18)=3.42$, $p<0.003$, $q<0.015$), and trended toward significance after the correction ($q<0.1$) for the Practice run (PR: $t(18)=2.19$, $p<0.042$, $q<0.069$) and the Transfer run (TR: $t(18)=2.52$, $p<0.021$, $q<0.053$). The effect sizes for these three runs were 0.78, 0.50, and 0.58, respectively. When the individual BOLD activity levels were averaged across the four nf runs (PR, R1, R2, R3), the group mean was significant ($t(18)=3.18$, $p<0.005$) with the effect size of 0.73. There was no significant difference in the LA activity levels between the Transfer run and Run 3 (TR vs R3: $t(18)=-1.66$, $p<0.114$). During the 2nd rtfMRI-nf session, the LA BOLD activity levels for the EG (Fig. 4A, middle) trended toward significance after the correction for the Practice run (PR: $t(18)=2.26$, $p<0.037$, $q<0.091$) and Run 2 (R2: $t(18)=2.56$, $p<0.020$, $q<0.091$). The effect sizes for these two runs were 0.52 and 0.59, respectively. The group mean for the individual activity levels averaged across the four nf runs in the 2nd session was significant ($t(18)=2.14$, $p<0.047$) with the effect size of 0.50. During the 3rd rtfMRI-nf session, the LA BOLD activity levels for the EG (Fig. 4A, right) were not significant. For the CG, the LA BOLD activity was not significant for any of the runs in the three nf sessions (Fig. 4B).

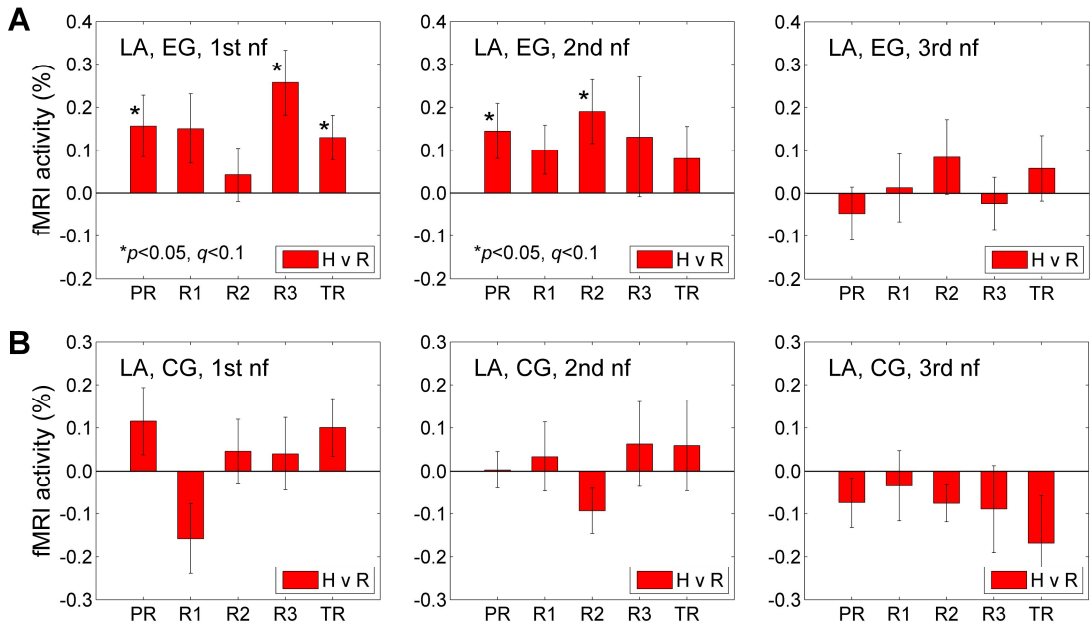

**Fig. 4.** BOLD activity of the left amygdala during the Happy Memories conditions in the three rtfMRI-nf sessions. **A)** Average fMRI percent signal changes for the left amygdala ROI (LA, Fig. 2C) across the five task runs in the 1st (visit 4), 2nd (visit 5), and 3rd (visit 6) rtfMRI-nf sessions (Fig. 1A) for the experimental group (EG). Each bar represents a mean GLM-based fMRI percent signal change for the Happy Memories conditions with respect to the Rest baseline (H vs R) in a given run, averaged across the group. The error bars are standard errors of the means (sem) for the group averages. The experimental runs and condition blocks are depicted schematically in Fig. 1B. **B)** Corresponding average fMRI percent signal changes for the control group (CG).

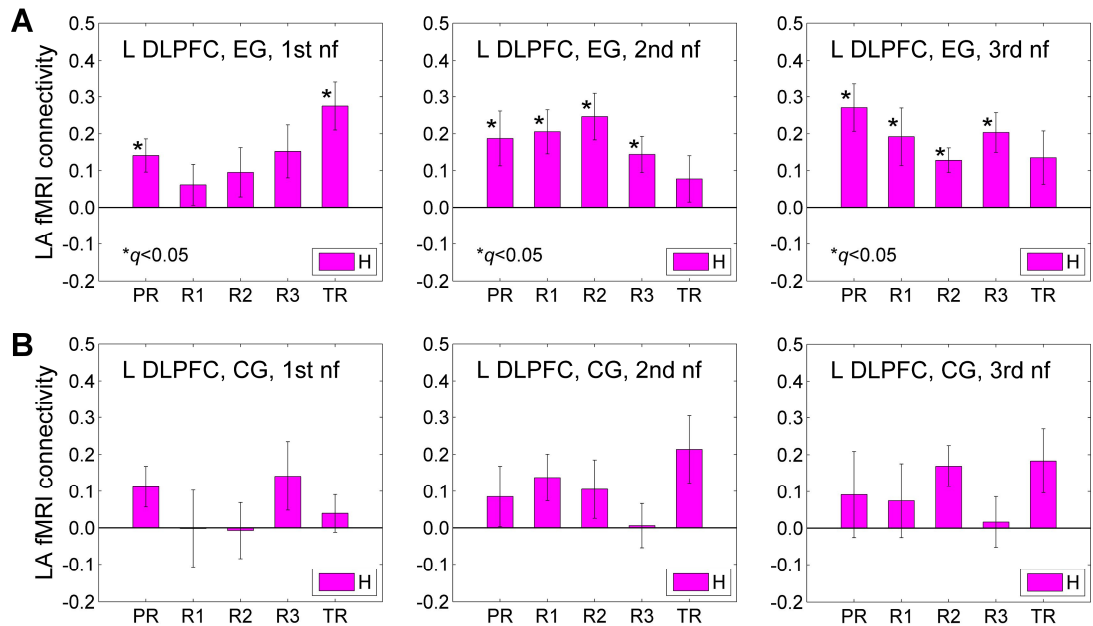

**Fig. 5.** fMRI functional connectivity between the left amygdala and the left dorsolateral prefrontal cortex (DLPFC) during the Happy Memories conditions in the three rtfMRI-nf sessions. **A)** Average fMRI connectivities of the LA seed ROI (Fig. 2C) with the left DLPFC ROI across the five task runs in the 1st (visit 4), 2nd (visit 5), and 3rd (visit 6) rtfMRI-nf sessions (Fig. 1A) for the experimental group (EG). The left DLPFC ROI is a 10-mm diameter ROI centered at (−45, 21, 24) in the Talairach space. This locus is shown in Fig. 10 below. Each bar represents a mean GLM-based fMRI connectivity strength ($z$-score) for the two ROIs during the Happy Memories conditions (H) in a given run, averaged across the group. The error bars are standard errors of the means (sem) for the group averages. **B)** Corresponding average fMRI connectivities for the control group (CG).

A 4 (Training: PR, R1, R2, R3) × 2 (Group: EG, CG) repeated measures ANOVA on the LA BOLD activity levels (Fig. 4) revealed a significant effect of the Group for the 1st rtfMRI-nf session ($F(1,28)=4.48$, $p<0.043$). The Training effect and the Training × Group interaction were not significant. Follow-up independent-samples $t$-tests showed the EG vs CG group difference for Run 1 that trended toward significance after correction (R1: $t(28)=2.51$, $p<0.018$, $q<0.090$). The group difference for Run 3 trended toward significance before correction (R3: $t(28)=1.84$, $p<0.076$, $q<0.190$). The effect sizes for these group differences were 0.95 and 0.70, respectively. For the 2nd and 3rd rtfMRI-nf sessions, the effects of the Group were not significant (2nd: $F(1,27)=2.12$, $p<0.157$; 3rd: $F(1,25)=0.90$, $p<0.351$).

The fMRI activation results for the LHIPS ROI (Fig. 2B), corresponding to the LA activation results in Fig. 4, are reported in *Supplementary material* (S2.2, Fig. S2).

### 3.3. Amygdala-DLPFC connectivity

Fig. 5 illustrates fMRI functional connectivity between the LA and the left DLPFC during the Happy Memories conditions (H) across the three rtfMRI-nf sessions. The results are for those participants who completed the study (EG: $n=15$, CG: $n=8$). The left DLPFC ROI was selected as a 10-mm diameter sphere centered at (−45, 21, 24) in the Talairach space. The changes in fMRI functional connectivity of this DLPFC region with the LA seed ROI (Fig. 2C) between the initial and final rtfMRI-nf sessions showed significant inverse correlation with the corresponding changes in PTSD severity for the EG, as described below (Section 3.6, Figs. 10, 11).

The results in Fig. 5A show significant fMRI connectivity (FDR $q<0.05$) between the LA and the left DLPFC for many runs during the 1st, 2nd, and 3rd nf sessions for the EG. Notably, the fMRI connectivities averaged across the four nf runs (PR, R1, R2, R3) in the 2nd and 3rd nf sessions were significantly higher than the average fMRI connectivity across the same runs in the 1st nf session (2nd vs 1st: $t(14)=2.71$, $p<0.017$, $d=0.70$; 3rd vs 1st: $t(14)=2.55$, $p<0.023$, $d=0.66$). For the CG, the LA-DLPFC connectivity was not significant for any of the runs in the three nf sessions (Fig. 5B). Average connectivities across the four nf runs did not differ between the sessions for the CG (2nd vs 1st: $t(7)=0.327$, $p<0.754$, $d=0.12$; 3rd vs 1st: $t(7)=0.334$, $p<0.749$, $d=0.12$). Average connectivity changes between the sessions were higher for the EG than for the CG, but not significantly (3rd vs 1st, EG vs CG: $t(21)=0.795$, $p<0.435$, $d=0.35$).

A 4 (Training: PR, R1, R2, R3) × 2 (Group: EG, CG) repeated measures ANOVA on the LA-DLPFC connectivity levels (Fig. 5) revealed a nonsignificant effect of the Group for the 1st rtfMRI-nf session ($F(1,21)=0.791$, $p<0.384$). Importantly, the Group effects for the 2nd and 3rd nf sessions were significant (2nd: $F(1,21)=4.39$, $p<0.049$; 3rd: $F(1,21)=4.55$, $p<0.045$).

### 3.4. Amygdala connectivity during Practice run

Fig. 6 exhibits whole-brain group statistical maps of the correlation between the LA fMRI connectivity during the rtfMRI-nf task in the Practice run of the 1st rtfMRI-nf session and the initial CAPS ratings for the EG. Data from $n=19$ EG participants were included in the analysis. (Results for one EG participant, whose initial CAPS rating was much higher, CAPS=95, than for the rest of the EG subjects, were excluded from the analysis to avoid biasing the group results). The group statistical maps in Fig. 6 were thresholded at $t=\pm 2.95$ (uncorr. $p<0.01$) and clusters containing at least 75 voxels (FWE corr. $p<0.05$) are shown in the figure. The cluster properties are described in Table 3. The results in Fig. 6 and Table 3 demonstrate that, at the beginning of the training, fMRI connectivity with the LA showed *negative* correlations with PTSD severity for many prefrontal brain regions, particularly the LOFC, the MOFC, the rACC, and the DLPFC. For the CG, the correlation results for the Practice run of the 1st nf session were similar to those for the EG in Fig. 6.

The negative correlation effects mapped in Fig. 6 are illustrated in Fig. 7. Note that several regions exhibited *positive* correlations between their fMRI connectivity with the LA and the initial CAPS ratings, but the corresponding clusters were not large enough to survive the whole-brain FWE correction. For example: the left caudate at (−19, −25, 20) ($t=5.21$, 62 voxels), the right mediodorsal nucleus (MD) of the thalamus at (3, −16, 15) ($t=5.46$, 26 voxels), the right precuneus (PCun, BA 39) at (27, −57, 31) ($t=4.24$, 22 voxels), the left precuneus (BA 7) at (−25, −61, 31) ($t=4.24$, 22 voxels). The positive correlation effect for the right precuneus is also illustrated in Fig. 7.

### 3.5. Amygdala connectivity enhancement across runs

Fig. 8 shows results of the whole-brain statistical group analysis of the correlation between the LA fMRI connectivity slope (FCS) across Happy Memories conditions in the four rtfMRI-nf runs (Fig. 3B) during the 1st rtfMRI-nf session and the initial CAPS ratings. The results are for the same EG participants ($n=19$) as in Figs. 6 and 7. The maps in Fig. 8 were thresholded at $t=\pm 2.95$ (uncorr. $p<0.01$) and clusters containing at least 81 voxels (FWE corr. $p<0.025$) are shown in the figure. The cluster properties are specified in Table 4. The table also includes statistical results for the mean FCS effect, thresholded and clustered the same way (FWE corr. $p<0.025$, to account for testing the two effects). The mean FCS effect was obtained in the same group analysis and corresponds to the mean values of the covariates (CAPS ratings, HDRS ratings, LA FCS for central WM).

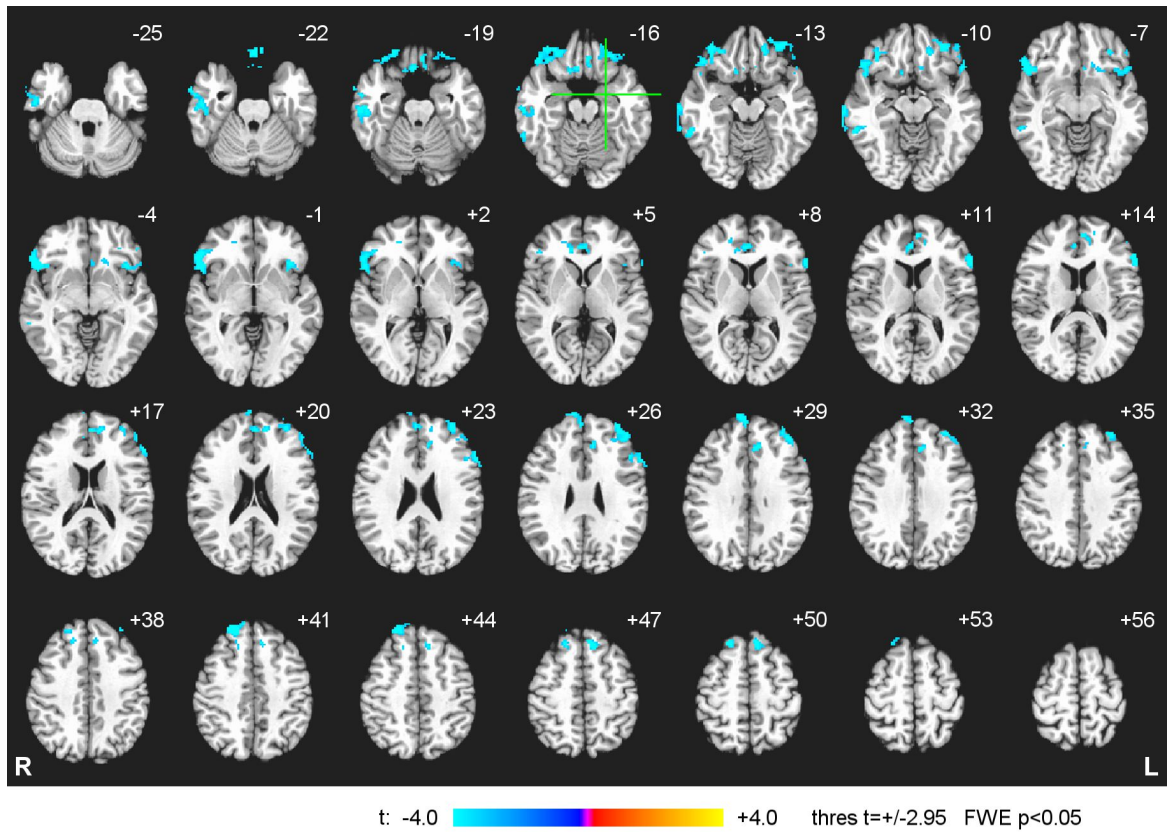

**Fig. 6.** Statistical maps of the correlation between the left amygdala fMRI connectivity during the Happy Memories conditions in the Practice run of the 1st rtfMRI-nf session and the initial PTSD severity for the experimental group (EG). The correlation is a voxel-wise partial correlation with the initial CAPS ratings controlled for comorbid depression severity (HDRS) ratings and average individual LA connectivity with central white matter. The maps are FWE corrected and projected onto the standard anatomical template TT_N27 in the Talairach space, with 3 mm separation between axial slices. The number adjacent to each slice indicates the $z$ coordinate in mm. The left hemisphere (L) is to the reader's right. The green crosshairs mark the center of the LA target ROI. Peak $t$-statistics values for the correlation effect and the corresponding cluster properties are specified in Table 3.

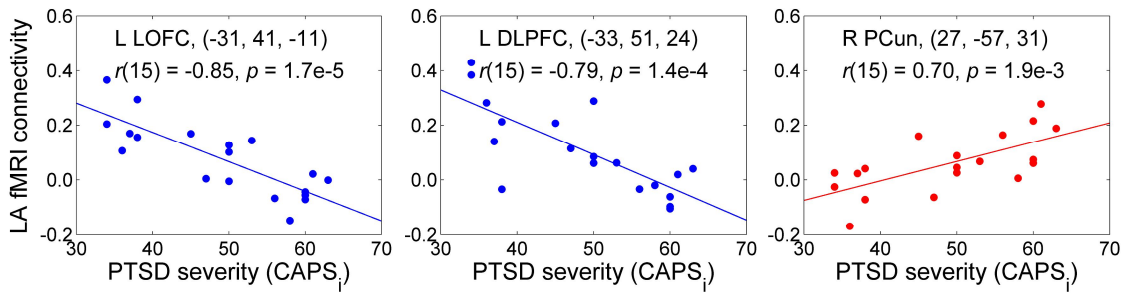

**Fig. 7.** Illustration of the correlation effects between the LA fMRI connectivity during the first Practice run and the initial PTSD severity, exhibited in Fig. 6. Each plot shows an average correlation effect for a 10-mm diameter ROI centered at a specified location. The correlation is a partial correlation with the initial CAPS ratings controlled for HDRS ratings and average individual LA connectivity with central white matter ($n=19$, $df=15$). The results for the left LOFC and the left DLPFC correspond to those reported in Fig. 6 and Table 3. The result for the right PCun is included to illustrate existence of positive correlations between the LA fMRI connectivity and PTSD severity (see text). Abbreviations: LOFC – lateral orbitofrontal cortex, DLPFC – dorsolateral prefrontal cortex, PCun – precuneus.

The results in Fig. 8 and Table 4 demonstrate that the fMRI connectivity enhancement with the LA during the training exhibited *positive* correlations with the initial PTSD severity for several prefrontal regions, including the LOFC and the DLPFC. The left DLPFC also showed a significant fMRI connectivity enhancement with the LA that was independent of the CAPS and HDRS variability (the mean FCS, Table 4). Note that the brain regions in

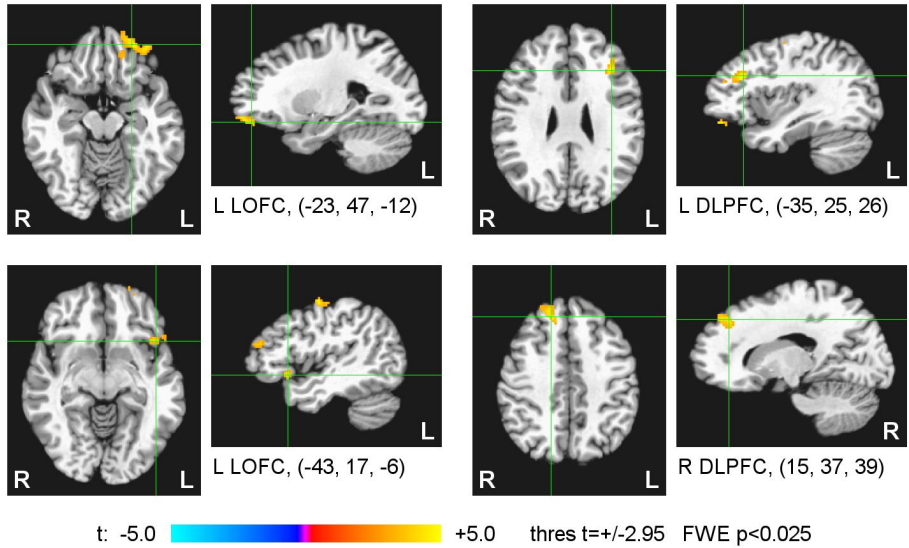

**Fig. 8.** Statistical maps of the correlation between the left amygdala fMRI connectivity slope (FCS) during the 1st rtfMRI-nf session and the initial PTSD severity for the experimental group (EG). The FCS is defined in Fig. 3B. The correlation is a voxel-wise partial correlation with the initial CAPS ratings controlled for comorbid depression severity (HDRS) ratings and average individual LA FCS for central white matter. The maps are FWE corrected. The green crosshairs mark the statistical peak locations, with their Talairach coordinates specified underneath. Peak $t$-statistics values and the corresponding cluster properties are described in Table 4. Abbreviations: LOFC – lateral orbitofrontal cortex, DLPFC – dorsolateral prefrontal cortex.

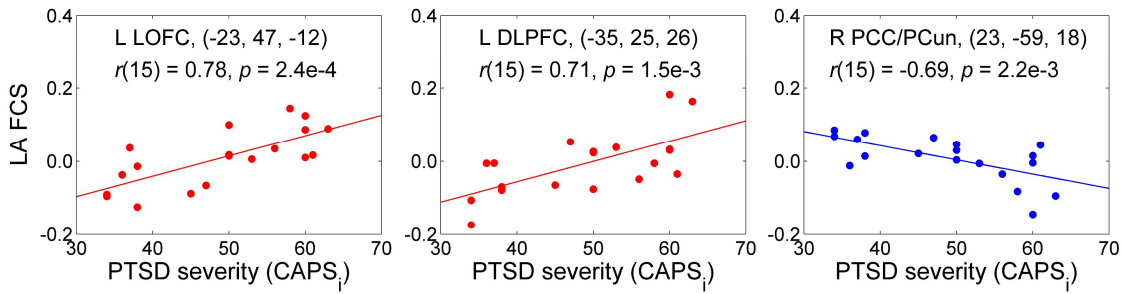

**Fig. 9.** Illustration of the correlation effects between the LA FCS and the initial PTSD severity, exhibited in Fig. 8. Each plot shows an average correlation effect for a 10-mm diameter ROI centered at a specified location. The correlation is a partial correlation with the initial CAPS ratings controlled for HDRS ratings and average individual LA FCS for central white matter ($n$=19, $df$=15). The results for the left LOFC and the left DLPFC correspond to those reported in Fig. 8 and Table 4. The result for the right posterior cingulate/precuneus (PCC/PCun) is included to illustrate existence of negative correlations between the LA FCS and PTSD severity (see text).

Fig. 8 and Table 4 have predominantly left lateralization. For the CG, no significant positive FCS vs CAPS correlations or mean FCS effects were found within the prefrontal cortex.

The positive correlation effects mapped in Fig. 8 are illustrated in Fig. 9. Note that several other regions that showed negative correlations in Figs. 6, 7 exhibited positive correlations between the FCS and CAPS, but the corresponding clusters did not survive the whole-brain FWE correction. For example: the right LOFC at (55, 25, −1) ($t$=5.22, 68 voxels), the left inferior temporal gyrus (ITG, BA 20) at (−48, −22, −20) ($t$=3.93, 45 voxels), the left MOFC at (−1, 39, −15) ($t$=3.99, 25 voxels). Furthermore, some regions showed *negative* correlations between the FCS and CAPS, e.g. the right posterior cingulate/precuneus (BA 31) at (23, −59, 18) ($t$=−5.80, 39 voxels). The negative correlation effect for this region is also illustrated in Fig. 9.

### 3.6. Amygdala connectivity changes between sessions

Fig. 10 shows results of the whole-brain statistical group analysis of the correlation between the average LA fMRI connectivity changes between the final (3rd) and initial (1st) rtfMRI-nf sessions and the corresponding changes (final vs initial) in the CAPS ratings. The results are for the EG participants who completed the study, including the final CAPS assessment ($n$=15). The maps in Fig. 10 were thresholded at $t$=±3.11 (uncorr. $p$<0.01) and clusters containing at least 81 voxels (FWE corr. $p$<0.025) are shown in the figure. The cluster properties are described in Table 5. The table also includes statistics for the mean LA fMRI connectivity changes between the sessions, thresholded and clustered the same way. The mean fMRI connectivity changes were obtained in the same group analysis and correspond to the mean values of the covariates (changes in CAPS ratings, changes in HDRS ratings, changes in LA fMRI connectivity for central WM).

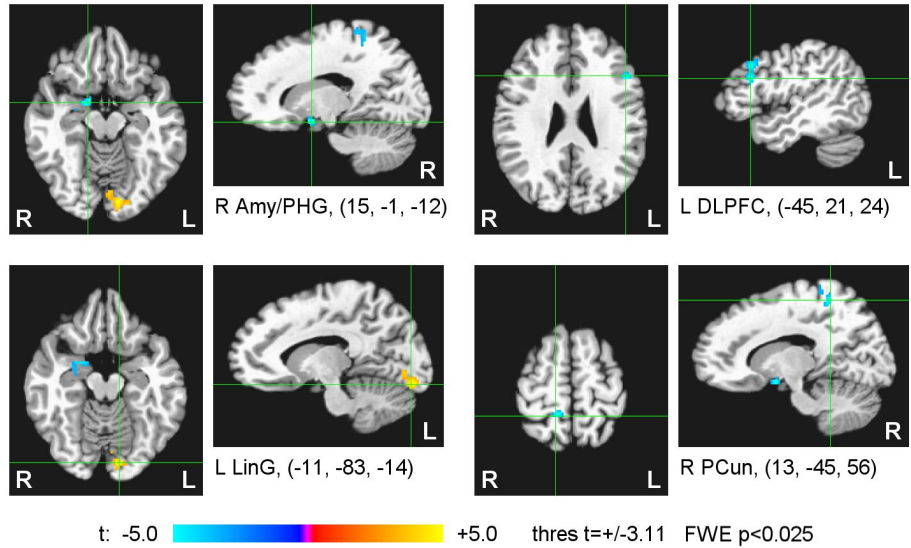

**Fig. 10.** Statistical maps of the correlation of the average left amygdala fMRI connectivity changes between the final (3rd) and initial (1st) rtfMRI-nf sessions and the corresponding changes in PTSD severity for the experimental group (EG). The LA fMRI connectivity values were averaged across Happy Memories conditions in the four rtfMRI-nf runs for each session. The correlation is a voxel-wise partial correlation with the changes in the CAPS ratings (final vs initial) controlled for corresponding changes in comorbid depression severity (HDRS) ratings and changes in average individual LA connectivity for central white matter. The maps are FWE corrected. The green crosshairs mark the statistical peak locations, with their Talairach coordinates specified underneath. Peak *t*-statistics values and the corresponding cluster properties are described in Table 5. Abbreviations: Amy – amygdala, PHG – parahippocampal gyrus, DLPFC – dorsolateral prefrontal cortex, LinG – lingual gyrus, PCun – precuneus.

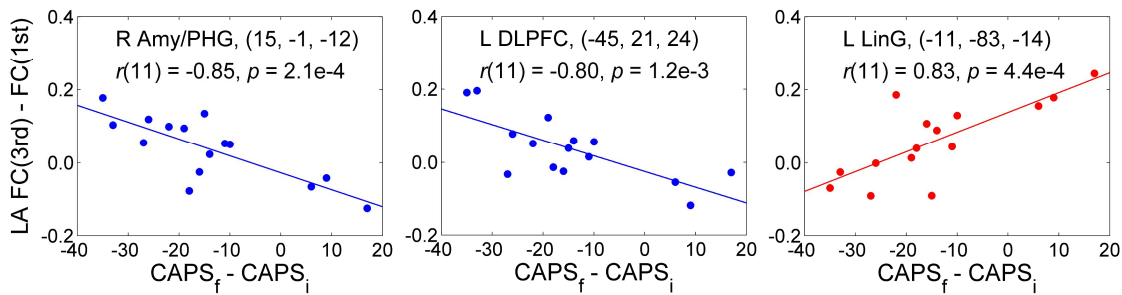

**Fig. 11.** Illustration of the correlation effects between the average LA fMRI connectivity changes and the corresponding PTSD severity changes for the EG, exhibited in Fig. 10. Each plot shows an average correlation effect for a 10-mm diameter ROI centered at a specified location. The correlation is a partial correlation with the changes in CAPS ratings (final vs initial) controlled for corresponding changes in HDRS ratings and changes in average individual LA connectivity for central white matter ($n=15$, $df=11$).

The results in Fig. 10 and Table 5 demonstrate that the LA fMRI connectivity changes between the two sessions exhibited *negative* correlations with the corresponding PTSD severity changes for the right amygdala/PHG, the left DLPFC, and the right superior precuneus. The right amygdala/PHG cluster has the statistical peak at (15, −1, −12) in BA 34 and the center of mass at (26, −2, −21) in the right amygdala/uncus. The left DLPFC cluster has the statistical peak at (−45, 21, 24) in BA 46 and the center of mass at (−46, 19, 32) in BA 9. According to Fig. 10 and Table 5, the LA fMRI connectivity changes for the lingual gyrus (BA 18) exhibited *positive* correlations with the PTSD severity changes. The mean fMRI connectivity changes between the sessions were positive (Table 5).

The correlation effects mapped in Fig. 10 are illustrated in Fig. 11. Note that some other regions also exhibited negative correlations between their LA fMRI connectivity changes and the corresponding CAPS changes, but the clusters were not large enough to survive the whole-brain FWE correction. For example: the left superior temporal gyrus (BA 38, temporal pole) at (−25, 19, −32) ($t=-4.90$, 72 voxels), the left LOFC (BA 47) at (−19, 21, −14) ($t=-5.04$, 45 voxels).

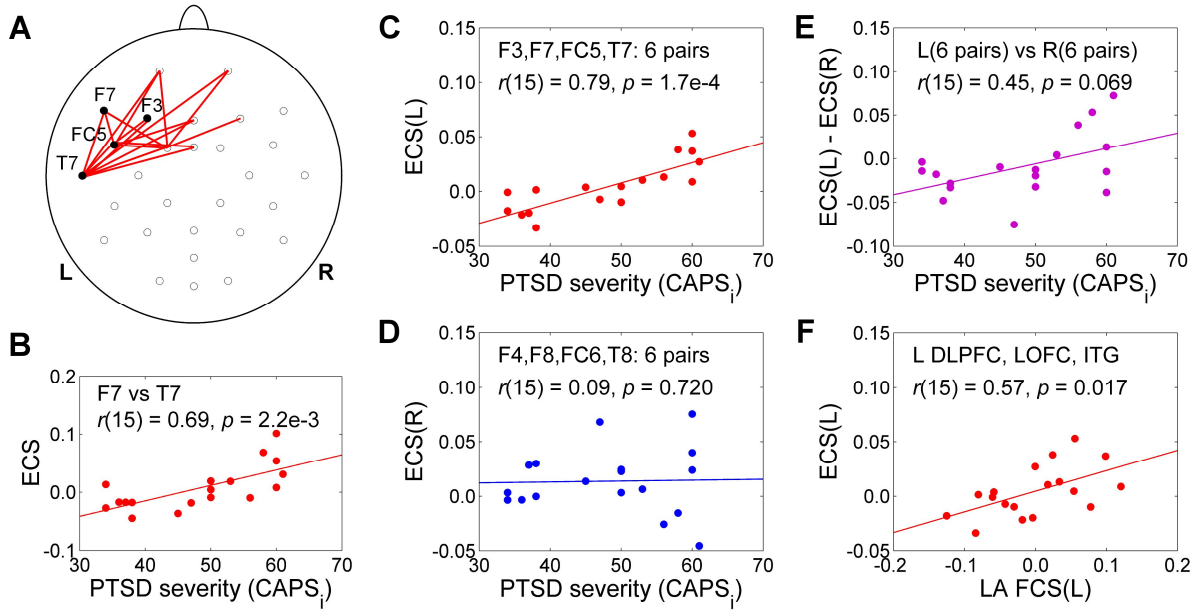

**Fig. 12.** Correlation between the EEG coherence slope (ECS) in the upper alpha band during the 1st rtfMRI-nf session and the initial PTSD severity for the experimental group (EG). The ECS is defined in Fig. 3C. The correlation for each EEG channel pair (or average across multiple channel pairs) is a partial correlation with the initial CAPS ratings controlled for comorbid depression severity (HDRS) ratings ($n=18$, $df=15$). **A)** Red segments denote EEG channel pairs for which the ECS vs CAPS correlations are positive ($r(15)>0$, $p<0.05$, uncorr.). **B)** Example of such correlation for one channel pair. **C)** Significant correlation between the ECS(L), i.e. the average ECS for 6 pairs of prefronto-temporal EEG channels on the left, and the CAPS ratings. **D)** Lack of correlation between the ECS(R), i.e. the average ECS for 6 pairs of corresponding prefronto-temporal EEG channels on the right, and the CAPS ratings. **E)** Correlation between the average ECS laterality, ECS(L)−ECS(R), and the CAPS ratings. **F)** Correlation between the average LA fMRI connectivity slope (FCS) for three ROIs in the prefronto-temporal regions on the left and the ECS(L). Abbreviations: DLPFC – dorsolateral prefrontal cortex, LOFC – lateral orbitofrontal cortex, ITG – inferior temporal gyrus.

Changes in LA FCS between the initial and final rtfMRI-nf sessions and their correlations with the corresponding changes in PTSD severity are examined in *Supplementary material* (S2.3, Figs. S3, S4).

### 3.7. EEG coherence enhancement across runs

Fig. 12 demonstrates correlations between the EEG coherence slope (ECS) for the upper alpha EEG band across the four nf runs (Fig. 3C) in the 1st rtfMRI-nf session and the initial CAPS ratings for the EG. The EEG recordings were conducted simultaneously with fMRI (Fig. 2D). Data from $n=18$ EG participants were included in the ECS vs CAPS correlation analysis. (One EG participant's results were excluded because of the very high initial CAPS rating as explained above; another participant's data were excluded due to excessive EEG-fMRI artifacts). According to Fig. 12A, the ECS exhibited positive partial correlations ($r(15)>0$, $p<0.05$, uncorr.) with the initial CAPS ratings for many EEG channel pairs, particularly those involving prefrontal (F3, F7, FC5) and temporal (T7) EEG channels on the left. (Negative correlations, $r(15)<0$, did not reach the $p<0.05$ statistical threshold). The ECS vs CAPS correlation effect is illustrated in Fig. 12B for one channel pair. Similar to our previous work (Zotev et al., 2016), we defined average ECS for six pairs of prefronto-temporal EEG channels on the left, ECS(L), and for six corresponding channel pairs on the right, ECS(R), as detailed in Fig. 12C,D. The ECS(L) demonstrated a significant *positive* correlation with the initial CAPS ratings (Fig. 12C). The average individual ECS laterality, ECS(L)−ECS(R), showed a positive correlation with CAPS that trended toward significance (Fig. 12E).

Correlation between the ECS and FCS metrics is illustrated in Fig. 12F. In the figure, the LA FCS(L) is the average FCS between the LA and three 10-mm diameter ROIs, centered at (−35, 25, 26) in the left DLPFC, at (−43, 17, −6) in the left LOFC, and at (−48, −22, −20) in the left ITG. These loci were reported above (Section 3.5). Note that these ROIs are located approximately underneath EEG channels F3, F7, and T7, respectively. The correlation in Fig. 12F is controlled for the average FCS for central WM, but not for CAPS or HDRS. According to Fig. 12F, there is a significant positive correlation between the average LA FCS for these three regions and the average ECS for the EEG channels above them. This correlation is mediated by PTSD severity.

## 4. Discussion

After three rtfMRI-nf emotion regulation training sessions, the EG participants who completed the study ($n=15$) showed a significant reduction in PTSD severity with large effect size (CAPS rating decrease: $p<0.0024$, $d=-0.95$) (Section 3.1, Table 2). This total CAPS score reduction was predominantly due to reduction in avoidance symptoms ($p<0.002$, $d=-0.98$), with moderately significant reduction in hyperarousal symptoms ($p<0.024$, $d=-0.65$) (Table 2). However, the CG participants who completed the study ($n=8$) also showed a reduction in PTSD severity, though non-significant and with 1.5 times smaller effect size ($p<0.124$, $d=-0.62$) (Table 2). No statistically significant difference in PTSD severity changes between the EG and CG could be demonstrated in the present work.

80% of the EG participants who completed the study achieved clinically meaningful reductions in PTSD severity (by 10 CAPS points or more) (Fig. S1). The 10 CAPS point threshold had been used in previous PTSD research (e.g. Gerin et al., 2016; Krystal et al., 2011; Steenkamp et al., 2015). The 80% symptom improvement rate is higher than that in clinical trials (50-70%) in which PTSD veterans underwent cognitive processing therapy or prolonged exposure therapy (Steenkamp et al., 2015). It is consistent with results of the pilot study on amygdala rtfMRI-nf by Gerin et al., 2016, in which two PTSD veterans out of three showed clinically meaningful PTSD symptom improvements. However, 38% of the CG participants who completed the study also demonstrated clinically meaningful reductions in PTSD severity (Fig. S1). This symptom improvement in the CG can be attributed to i) positive emotion induction during the rtfMRI-nf task, and ii) beneficial psychological effects of regular interactions with the clinical assessment personnel in the course of the study.

Compared to the reduction in PTSD avoidance symptoms ($p<0.002$, $d=-0.98$), the reduction in comorbid depression severity for the EG was even more significant with larger effect size (HDRS rating decrease: $p<0.0004$, $d=-1.19$) (Table 2). This finding is consistent with the beneficial effects of the amygdala rtfMRI-nf in MDD patients (Young et al., 2017). Previous studies have linked depression to deficient approach motivation (e.g. Bruder et al., 2017; Henriques & Davidson, 2000; McFarland et al., 2006). Thus, the reduction in depression severity could conceivably be associated with an enhancement in approach motivation (though approach tendencies were not directly assessed in the present study). This reasoning suggests that the strongest neuropsychological effects of the rtfMRI-nf training in PTSD veterans may occur along the approach-avoidance (motivational) dimension. Such interpretation is consistent with a stronger engagement of the EF/ER system during performance of the rtfMRI-nf task, which represents a goal-oriented behavior as we discussed previously (Zotev et al., 2016).

During the 1st rtfMRI-nf training session, the EG participants were able to successfully upregulate the LA BOLD activity (Section 3.2, Fig. 4A, left). The mean LA activity for the individual activity levels averaged across the four nf runs was significant with medium effect size ($p<0.005$, $d=0.73$). The highest LA activation was observed for Run 3 ($p<0.003$, $d=0.78$), suggesting that the participants gradually learned to upregulate the LA activity as the training progressed. The effect size $d=0.73$ for the average results across the four nf runs is lower than the $d=0.87$ effect size computed in the same way for MDD patients ($n=13$) who performed the same procedure in our previous study (Zotev et al., 2016). Therefore, a larger group size is needed in the case of PTSD patients to reach a comparable statistical power (Desmond & Glover, 2002). Importantly, the LA BOLD activity levels averaged across the four nf runs in the 1st rtfMRI-nf session were significantly higher for the EG than for the CG ($p<0.043$).

During the 2nd rtfMRI-nf session, the EG participants were also able to upregulate the LA BOLD activity (Section 3.2, Fig. 4A, middle). The mean activity for the individual LA BOLD activity levels averaged across the four nf runs was significant with medium effect size ($p<0.047$, $d=0.50$). Unfortunately, no significant upregulation of the LA BOLD activity was observed during the 3rd rtfMRI-nf session (Fig. 4A, right). Examination of the POMS and VAS mood rating changes (not included here) showed that the EG participants *were* able to induce happy emotion (subjectively rated) during the 3rd rtfMRI-nf session. Therefore, we tentatively attribute the diminished rtfMRI-nf performance during the 3rd session to an insufficient effort put by the participants into upregulation of the rtfMRI-nf signal. We believe that such drop in rtfMRI-nf performance can be prevented in future studies through the following measures: i) careful evaluation of each participant's performance and personal experiences after each nf session; ii) development of more effective personalized mental strategies and performance encouragement for the next nf session.

Notably, we observed increased fMRI functional connectivity between the LA and the selected left DLPFC region during the rtfMRI-nf task across the three rtfMRI-nf sessions for the EG (Section 3.3, Fig. 5A). The mean LA-DLPFC connectivity strength for the individual connectivities averaged across the four nf runs was significantly higher ($p<0.023$, $d=0.66$) during the 3rd rtfMRI-nf session (Fig. 5A, right) than during the 1st session (Fig. 5A, left) for the EG. It was also significantly higher for the EG than for the CG during both the 2nd ($p<0.049$) and the 3rd ($p<0.045$) nf sessions. This LA-DLPFC connectivity enhancement for the EG was observed despite the fact that the mean LA BOLD activity levels did not increase from session to session (Fig. 4A). This comparison suggests that a higher mean LA activity does not necessarily correspond to a stronger fMRI connectivity between the LA and the prefrontal regions involved in emotion regulation.

To characterize task-specific LA connectivity at the beginning of the training, we examined fMRI connectivity of the LA during the rtfMRI-nf task in the Practice run of the 1st rtfMRI-nf session (Section 3.4). During this run, the participants were exposed to the rtfMRI-nf for the first time and did not yet know how to effectively control the rtfMRI-nf signal. The results in Fig. 6 and Table 3 demonstrate negative correlations between the LA fMRI connectivity and the initial CAPS ratings for many prefrontal regions, including the LOFC (BA 47, 11), the MOFC (BA 11), the DLPFC (BA 9, 8), the VLPFC (BA 45), the medial frontopolar cortex (BA 9), and the rACC (BA 24). These results are consistent with the previously reported pattern of PFC hypoactivity in PTSD (e.g. Etkin & Wager, 2007; Lanius et al., 2006; Patel et al., 2012). At the same time, the LA connectivity with several brain regions, including the right mediodorsal nucleus of the thalamus (MD) and the bilateral precuneus (BA 39, 7), exhibited positive, though less significant, correlations with the initial CAPS ratings (Section 3.4, Fig. 7). Note that parietal regions, including the precuneus and the inferior parietal lobule, are known to be hyperactive in PTSD together with the amygdala (e.g. Etkin & Wager, 2007; Lanius et al., 2006; Patel et al., 2012). Our results suggest that fMRI connectivity between the amygdala and regions involved in autobiographical memory recall (MD, precuneus) is elevated in PTSD not only during recollection of traumatic events, but also during retrieval of happy autobiographical memories.

The main result of the present work is the observation of the significant *positive* correlations between the LA fMRI connectivity enhancement (FCS) for several PFC regions across nf runs in the 1st rtfMRI-nf session and the initial PTSD severity (Section 3.5, Fig. 8). This positive FCS vs CAPS correlation effect is observed for the left LOFC (BA 47, 11), the bilateral DLPFC (BA 9), and the left precentral gyrus (BA 4) for the EG (Fig. 8, Table 4). Positive, though less significant, correlation effects are also found for the right LOFC (BA 47), the left ITG (BA 20), and the left MOFC (BA 11) (Section 3.5). Such positive correlations indicate that the patients with more severe PTSD (higher initial CAPS ratings) showed *more positive* changes in the LA connectivity with these PFC regions as the rtfMRI-nf training progressed. For the right PCC/precuneus, the corresponding LA connectivity changes were more negative (Fig. 9). Therefore, the results in Figs. 8 and 9 demonstrate correction (at least partial) of the LA functional connectivity abnormalities specific to PTSD and evident in Figs. 6 and 7. Importantly, these results reveal the underlying *mechanism* of the beneficial effects of the amygdala rtfMRI-nf in PTSD: normalization of the amygdala-prefrontal functional connectivity deficiencies. Furthermore, the EG participants exhibited a significant mean FCS effect in the left DLPFC (BA 9) (Table 4). This effect indicates a significant fMRI connectivity enhancement between the LA and the left DLPFC across the four nf runs, independent of the PTSD and depression severity. This finding is generally consistent with the positive group-average fMRI connectivity changes between the amygdala and the DL/DMPFC during the rtfMRI-nf task reported by Nicholson et al., 2017.

Reduction in PTSD severity between the initial and final rtfMRI-nf sessions for the EG was associated with enhancement in LA functional connectivity with the right amygdala/PHG (BA 34) and the left DLPFC (BA 46/9) (Section 3.6, Fig. 10). The negative correlation effect for the right amygdala/PHG in Figs. 10, 11 suggests that the EG participants, who engaged the right amygdala together with the LA during the rtfMRI-nf task, showed a larger reduction in PTSD symptom severity. The left DLPFC region in Fig. 10 is spatially close to the left DLPFC region in Fig. 8. The positive FCS vs initial CAPS correlation effect for the left DLPFC in Figs. 8, 9 is consistent with the negative FC change vs CAPS change correlation effect for the left DLPFC in Figs. 10, 11. Both effects correspond to correction of the PTSD-specific functional connectivity deficiencies between the LA and the left DLPFC, leading to reduction in PTSD severity. A similar, but less significant, FC change vs CAPS change correlation effect is observed for the left LOFC (BA 47) (Section 3.6).

In contrast to the negative FC change vs CAPS change correlation for the left DLPFC in Figs. 10, 11, the corresponding correlation for the left lingual gyrus (BA 18) is positive. This region is involved in visual memory (e.g. Slotnick, 2004). Thus, reduction in PTSD severity is associated with diminished connectivity between the LA and the visual memory system during the rtfMRI-nf task. Furthermore, reduction in PTSD severity is associated with reduced FCS between the LA and the posterior nodes of the default mode network (DMN), including the right PCC and the right angular gyrus (Figs. S3, S4). These regions are also involved in episodic memory retrieval (e.g. Sestieri et al., 2011). The negative FCS vs initial CAPS correlation effect for the right PCC/precuneus in Fig. 9 is consistent with the positive FCS change vs CAPS change correlation effect for the right PCC in Figs. S3, S4. Therefore, the EG participants, who showed a stronger suppression of functional connectivity between the LA and the occipito-parietal regions involved in memory functions, achieved a greater reduction in PTSD symptom severity.

EEG recordings performed simultaneously with fMRI allowed us to investigate electrophysiological correlates of the rtfMRI-nf training (Section 3.7). We examined variations in EEG coherence, which is an EEG measure of functional connectivity, across the four nf runs in the 1st rtfMRI-nf session (Fig. 12). The average enhancement in upper alpha EEG coherence for the prefronto-temporal EEG channels on the left, ECS(L), significantly correlated with the initial PTSD severity for the EG (Fig. 12C). Note that this positive ECS vs CAPS correlation effect is related to the positive FCS vs CAPS correlation effect in Fig. 8. Indeed, four out of five clusters in Fig. 8 appear within the left PFC. Stronger functional connectivities between these PFC regions and the LA mean stronger functional connectivities among them, leading to stronger coherences for EEG signals measured above these regions. This connection between the ECS and FCS is illustrated in Fig. 12F and explained in Section 3.7. Therefore, the PTSD-specific enhancements in functional connectivity that accompany the rtfMRI-nf training can be independently observed in both fMRI and EEG domains.

Interestingly, the map of ECS vs CAPS correlation effects in the EEG channel space (Fig. 12A) is similar to the map of ECS vs HDRS correlation effects in our recent MDD study (Zotev et al., 2016). However, the average ECS laterality, ECS(L)−ECS(R), exhibited a less significant positive correlation with the initial CAPS ratings (Fig. 12E) than the corresponding ECS(L) (Fig. 12C). This trend is different from that observed in our MDD study, in which the ECS laterality showed a more significant correlation with the MDD patients' HDRS ratings than the ECS(L) (Zotev et al., 2016). This result suggests that hemispheric EEG asymmetry/laterality effects, as revealed during the rtfMRI-nf training, may be less pronounced in PTSD than in MDD.

The reported study has several limitations. First, the rtfMRI-nf procedure did not include any personalized trauma-related content. Inclusion of such content will require downregulation of the amygdala activity (Gerin et al., 2016; Nicholson et al., 2017). This approach will be explored in our future work. Second, the study design focused on the correction of emotion regulation deficiencies in PTSD, and did not explicitly target other symptoms of PTSD, such as reexperiencing or avoidance. Third, the rtfMRI-nf task involved upregulation of the amygdala activity to enhance positive emotion, while the amygdala is usually hyperactive during emotional processing in PTSD. Nevertheless, our experimental results suggest that it is the dynamic process of volitional modulation of the amygdala activity using rtfMRI-nf that enhances the amygdala-prefrontal functional connectivity and benefits PTSD patients.

## 5. Conclusion

Our study demonstrates that the rtfMRI-nf training of the amygdala activity to enhance emotion regulation ability may be beneficial to veterans with combat-related PTSD. Our fMRI and EEG results independently suggest that the rtfMRI-nf training has the potential to correct the amygdala-prefrontal functional connectivity deficiencies specific to PTSD. The most significant PTSD-specific enhancements in fMRI connectivity between the LA and the PFC are observed for the left DLPFC and the left LOFC regions, which are parts of the executive function and emotion regulation system. Because activities of these cortical regions can be probed using scalp EEG, a carefully designed EEG-nf procedure may complement the rtfMRI-nf of the amygdala.

## Acknowledgments

This work was supported by the W81XWH-12-1-0697 grant from the U.S. Department of Defense. The authors thank all staff members at the Laureate Institute for Brain Research. We are particularly grateful to Dr. Matthew Meyer, MD and Dr. William Yates, MD for conducting psychiatric interviews, Tim Collins, Lisa Kinyon, and Megan Cole for administering clinical interviews and assessments, Julie Owen, Julie Crawford, Leslie Walker, and Tressia Lewis for helping with MRI and EEG-fMRI scanning. We would like to thank Dr. Patrick Britz, Dr. Robert Störmer, Dr. Mario Bartolo, and Dr. Brett Bays of Brain Products, GmbH for their help and technical support.

Table 1. Main characteristics of the experimental and control groups. Numbers of participants who started the study (and finished the first rtfMRI-nf session) and those who completed the whole study are specified for each group. Initial PTSD and depression severity ratings were measured at the beginning of the study prior to the rtfMRI-nf training sessions. Final PTSD and depression severity ratings were determined at the end of the study. Abbreviations: CAPS – Clinician-Administered PTSD Scale; PCL-M – PTSD CheckList Military Version; HDRS – Hamilton Depression Rating Scale; MADRS – Montgomery-Asberg Depression Rating Scale.

|  | Experimental group (EG) | | Control group (CG) | |
|---|---|---|---|---|
| **Measure** | **Initial mean (SD)** | **Final mean (SD)** | **Initial mean (SD)** | **Final mean (SD)** |
| Participants | 20 | 15 | 11 | 8 |
| Age | 31.0 (5.7) | 30.8 (5.4) | 34.1 (8.5) | 36.8 (8.0) |
| CAPS | 51.3 (14.3) | 40.6 (18.5) | 57.0 (25.3) | 53.8 (23.9) |
| PCL-M | 44.7 (10.8) | 36.3 (11.9) | 47.2 (17.8) | 40.5 (19.6) |
| HDRS | 16.8 (5.9) | 11.1 (5.7) | 14.7 (9.0) | 11.1 (6.2) |
| MADRS | 20.7 (8.9) | 13.9 (10.0) | 17.1 (12.9) | 14.3 (9.4) |

Table 2. Changes in PTSD severity and comorbid depression severity ratings for participants who completed the study. Mean rating values at the beginning of the study (initial) and at the end of the study (final) are included, and their statistical changes (final vs initial) within each group are specified. Abbreviations: CAPS – Clinician-Administered PTSD Scale; HDRS – Hamilton Depression Rating Scale.

| Rating | Initial mean (SD) | Final mean (SD) | Effect size ($d$) | Change $t$-score# | Change $p$-value [$q$] |
|---|---|---|---|---|---|
| **Experimental group (EG, $n$=15)** | | | | | |
| CAPS | 54.9 (14.1) | 40.6 (18.5) | −0.95 | −3.69 | 0.0024 [0.004]* |
|    Reexperiencing symptoms | 11.1 (6.05) | 9.07 (6.88) | −0.40 | −1.56 | 0.142 [0.142] |
|    Avoidance symptoms | 21.8 (7.65) | 14.0 (9.73) | −0.98 | −3.78 | 0.0020 [0.004]* |
|    Hyperarousal symptoms | 22.0 (4.90) | 17.5 (7.41) | −0.65 | −2.54 | 0.024 [0.030]* |
| HDRS | 17.3 (6.70) | 11.1 (5.72) | −1.19 | −4.61 | 0.0004 [0.002]* |
| **Control group (CG, $n$=8)** | | | | | |
| CAPS | 62.3 (22.4) | 53.8 (23.9) | −0.62 | −1.75 | 0.124 [0.207] |
|    Reexperiencing symptoms | 15.8 (6.78) | 13.5 (8.93) | −0.50 | −1.41 | 0.203 [0.254] |
|    Avoidance symptoms | 24.6 (12.3) | 18.0 (14.0) | −0.69 | −1.95 | 0.092 [0.207] |
|    Hyperarousal symptoms | 21.9 (4.94) | 22.3 (4.83) | 0.10 | 0.30 | 0.776 [0.776] |
| HDRS | 16.8 (9.07) | 11.1 (6.24) | −0.81 | −2.28 | 0.056 [0.207] |

# $t(14)$ for the EG, $t(7)$ for the CG.
* FDR $q<0.05$ for the five tests.

**Table 3.** Correlation between the fMRI connectivity of the left amygdala during the Happy Memories conditions in the Practice run of the 1st rtfMRI-nf session and the initial PTSD severity (CAPS). Location of the point with the peak group *t*-score and the number of voxels are specified for each cluster obtained after FWE correction for multiple comparisons.

| Region | Laterality | x, y, z (mm) | t-score | Size (# voxels) |
|---|---|---|---|---|
| **Frontal Lobe** | | | | |
| Lateral orbitofrontal cortex (BA 47) | R | 51, 23, −8 | −9.50 | 646 |
| Dorsolateral prefrontal cortex (BA 9) | L | −33, 51, 24 | −5.45 | 306 |
| Lateral orbitofrontal cortex (BA 11) | L | −31, 41, −11 | −6.77 | 288 |
| Medial frontal polar cortex (BA 9) | R | 9, 61, 30 | −6.91 | 252 |
| Ventrolateral prefrontal cortex (BA 45) | L | −55, 21, 12 | −6.85 | 188 |
| Dorsolateral prefrontal cortex (BA 8) | R | 21, 43, 42 | −6.70 | 181 |
| Superior frontal gyrus (BA 8) | L | −7, 31, 48 | −5.62 | 158 |
| Medial orbitofrontal cortex (BA 11) | R | 1, 33, −22 | −5.55 | 125 |
| Medial orbitofrontal cortex (BA 11) | R | 5, 20, −20 | −7.34 | 84 |
| **Temporal Lobe** | | | | |
| Middle temporal gyrus (BA 21) | R | 67, −19, −12 | −5.98 | 325 |
| Middle temporal gyrus (BA 20) | R | 57, −43, −12 | −5.94 | 92 |
| **Limbic Lobe** | | | | |
| Anterior cingulate cortex (BA 24) | R | 3, 37, 9 | −5.76 | 127 |
| **Sub-lobar Regions** | | | | |
| Insula (BA 13) | L | −35, 21, 0 | −4.79 | 144 |

Notations: BA – Brodmann areas; L – left; R – right; x, y, z – Talairach coordinates;
FWE corrected $p<0.05$ (Size – cluster size, minimum 75 voxels for uncorr. $p<0.01$).

**Table 4.** Correlation between the left amygdala fMRI connectivity slope (FCS) during the 1st rtfMRI-nf session and the initial PTSD severity (CAPS). Statistics for the mean FCS effect, independent of the CAPS and HDRS variations, are also included. Location of the point with the peak group *t*-score and the number of voxels are specified for each cluster obtained after FWE correction for multiple comparisons for each of the two effects.

| Region | Laterality | x, y, z (mm) | t-score | Size (# voxels) |
|---|---|---|---|---|
| **FCS vs CAPS correlation** | | | | |
| Lateral orbitofrontal cortex (BA 11) | L | −23, 47, −12 | 5.19 | 219 |
| Dorsolateral prefrontal cortex (BA 9) | R | 15, 37, 39 | 4.65 | 116 |
| Dorsolateral prefrontal cortex (BA 9) | L | −35, 25, 26 | 5.45 | 110 |
| Lateral orbitofrontal cortex (BA 47) | L | −43, 17, −6 | 4.69 | 106 |
| Precentral gyrus (BA 4) | L | −49, −11, 50 | 5.76 | 84 |
| **Mean FCS** | | | | |
| Dorsolateral prefrontal cortex (BA 9) | L | −55, 9, 30 | 4.68 | 250 |
| Inferior temporal gyrus (BA 20) | L | −55, −37, −16 | 6.35 | 93 |
| Superior temporal gyrus (BA 22) | R | 55, −7, −2 | 4.64 | 81 |

Notations: BA – Brodmann areas; L – left; R – right; x, y, z – Talairach coordinates;
FWE corrected $p<0.025$ (Size – cluster size, minimum 81 voxels for uncorr. $p<0.01$).

**Table 5.** Correlation of the average left amygdala fMRI connectivity (FC) changes between the 3rd and 1st rtfMRI-nf sessions and the corresponding changes in PTSD severity (final vs initial CAPS). Statistics for the mean FC changes between the two sessions, independent of the CAPS and HDRS rating changes, are also included. Location of the point with the peak group *t*-score and the number of voxels are specified for each cluster obtained after FWE correction for multiple comparisons for each of the two effects.

| Region | Laterality | x, y, z (mm) | t-score | Size (# voxels) |
|---|---|---|---|---|
| **FC changes vs CAPS changes correlation** | | | | |
| Amygdala / Parahippocampal gyrus (BA 34) | R | 15, −1, −12 | −8.15 | 250 |
| Lingual gyrus (BA 18) | L | −11, −83, −14 | 6.99 | 133 |
| Precuneus (BA 7) | R | 13, −45, 56 | −6.38 | 105 |
| Dorsolateral prefrontal cortex (BA 46/9) | L | −45, 21, 24 | −6.67 | 100 |
| **Mean FC changes** | | | | |
| Declive (cerebellum) | L | −23, −75, −20 | 7.10 | 547 |
| Precuneus (BA 7) | R | 9, −49, 42 | 7.14 | 503 |
| Insula (BA 13) | R | 37, 11, 0 | 4.87 | 237 |
| Lingual gyrus (BA 18) | L | −9, −59, 4 | 6.42 | 224 |
| Middle occipital gyrus | L | −31, −75, 8 | 5.31 | 219 |
| Middle temporal gyrus (BA 20) | L | −51, −37, −12 | 5.72 | 149 |
| Superior parietal lobule (BA 5) | R | 21, −39, 60 | 5.55 | 123 |
| Cuneus (BA 17) | R | 19, −81, 12 | 7.54 | 113 |
| Amygdala / Parahippocampal gyrus (BA 34) | L | −21, −1, −12 | 6.32 | 111 |
| Middle temporal gyrus (BA 19) | R | 41, −67, 14 | 6.52 | 106 |
| Superior temporal gyrus (BA 22) | R | 47, −23, 0 | 5.75 | 102 |
| Superior temporal gyrus (BA 22) | L | −51, 9, −6 | 4.64 | 96 |
| Postcentral gyrus (BA 2) | R | 33, −27, 38 | 4.54 | 87 |

Notations: BA – Brodmann areas; L – left; R – right; x, y, z – Talairach coordinates;
FWE corrected $p<0.025$ (Size – cluster size, minimum 81 voxels for uncorr. $p<0.01$).

# SUPPLEMENTARY MATERIAL

*S1.1. fMRI data analysis*

Offline analysis of the fMRI data was performed in AFNI (Cox, 1996; Cox & Hyde, 1997). Pre-processing of single-subject fMRI data included despiking using the 3dDespike AFNI program and correction of cardio-respiratory artifacts using the AFNI implementation of the RETROICOR method (Glover et al., 2000). Further fMRI pre-processing involved slice timing correction and volume registration of all EPI volumes acquired in the experiment using the 3dvolreg AFNI program with two-pass registration. The last volume of the short EPI dataset acquired immediately after the high-resolution anatomical (MPRAGE) brain image was used as the registration base.

The fMRI activation analysis was performed using the standard general linear model (GLM) approach. It was conducted for each of the five task fMRI runs (Fig. 1B) using the 3dDeconvolve AFNI program. The GLM model included two block-design stimulus condition terms, Happy Memories and Count (Fig. 1B), represented by the standard block-stimulus regressors in AFNI. A general linear test term was included to compute the Happy vs Count contrast. Nuisance covariates included the six fMRI motion parameters and five polynomial terms for modeling the baseline. To further reduce effects of residual motion artifacts, the fMRI data and motion parameters were lowpass Fourier filtered at 0.1 Hz prior to the GLM analysis. GLM $\beta$ coefficients were computed for each voxel, and average percent signal changes for Happy vs Rest, Count vs Rest, and Happy vs Count contrasts were obtained by dividing the corresponding $\beta$ values (×100%) by the $\beta$ value for the constant baseline term. The resulting fMRI percent signal change maps for each run were transformed to the Talairach space by means of the @auto_tlrc AFNI program using each subject's high-resolution anatomical brain image as the template.

Average individual BOLD activity levels for the left and right amygdala were computed in the offline analysis for the LA and RA ROIs, exhibited in Fig. 2C. The ROIs were defined anatomically as specified in the AFNI implementation of the Talairach-Tournoux brain atlas. The voxel-wise fMRI percent signal change data from the GLM analysis, transformed to the Talairach space, were averaged within the LA and RA ROIs and used as GLM-based measures of these regions' BOLD activities. Similarly, average individual BOLD activity levels were computed for the LHIPS target ROI in the Talairach space (Fig. 2B).

*S1.2. EEG data analysis*

Removal of MR and cardioballistic (CB) artifacts was based on the average artifact subtraction method implemented in BrainVision Analyzer 2.1 (Brain Products, GmbH). The MR artifact template was defined using MRI slice markers recorded with the EEG data. After the MR artifact removal, the EEG data were bandpass filtered between 0.5 and 80 Hz (48 dB/octave) and downsampled to 250 S/s sampling rate (4 ms interval). The fMRI slice selection frequency (17 Hz) and its harmonics were removed by band rejection filtering. The CB artifact template was determined from the cardiac waveform recorded by the ECG channel, and the CB artifact to be subtracted was defined, for each channel, by a moving average over 21 cardiac periods. Intervals with strong motion artifacts were not included in the CB correction.

Following the MR and CB artifact removal, the EEG data from the five task runs (Fig. 1B) were concatenated to form a single dataset. The data were carefully examined, and intervals exhibiting significant motion or instrumental artifacts ("bad intervals") were excluded from the analysis. Channel Cz was selected as a new reference, and FCz was restored as a regular channel.

An independent component analysis (ICA) was performed over the entire dataset with exclusion of the bad intervals. This approach ensured that independent components (ICs) corresponding to various artifacts were identified and removed in a consistent manner across all five runs. Channels TP9 and TP10 were excluded from the ICA and further analysis, because their signals are very sensitive to head and jaw movements, producing large artifacts. The Infomax ICA algorithm (Bell & Sejnowski, 1995), implemented in BrainVision Analyzer 2.1, was applied to the data from 29 EEG channels and yielded 29 ICs. Time courses, spectra, topographies, and kurtosis values of all the ICs were carefully analyzed (see e.g. McMenamin et al., 2010 and supplement therein) to identify various artifacts, as well as EEG signals of neuronal origin, with particular attention to the alpha and theta EEG bands. After all the ICs had been classified, an inverse ICA transform was applied to remove the identified artifacts from the EEG data. Following the ICA-based artifact removal, the EEG data were lowpass filtered at 40 Hz (48 dB/octave) and downsampled to 125 S/s (8 ms interval). Because many artifacts had been already removed using the ICA, the data were examined again, and new bad intervals were defined to exclude remaining artifacts.

*S2.1. Psychological measures and drop-out*

Fig. S1 illustrates individual changes in PTSD severity for the participants who completed the study. The corresponding mean CAPS ratings and statistics of their changes are reported in Table 2.

Mean initial PTSD and depression severity ratings for the participants who completed the whole study and for those who dropped out before completion are reported in

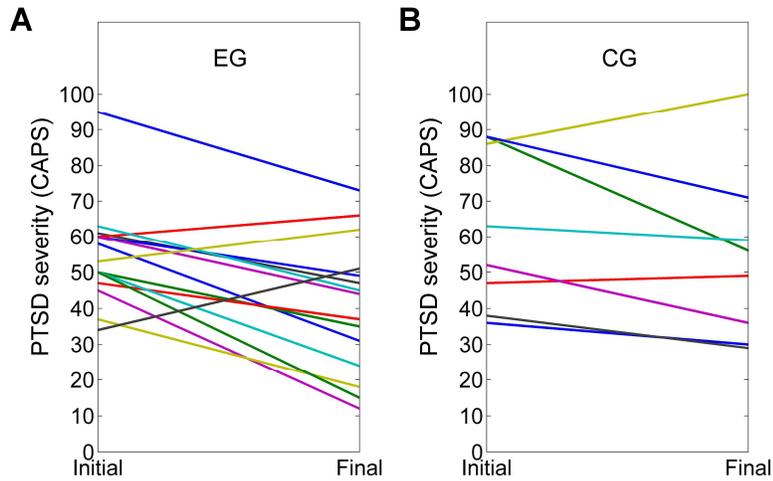

**Fig. S1.** Individual changes in PTSD severity for participants who completed the study. The initial and final PTSD severity (CAPS) ratings were determined at the beginning and at the end of the study, respectively. Each colored segment corresponds to one participant. **A)** Experimental group (EG, *n*=15). **B)** Control group (CG, *n*=8). Abbreviation: CAPS – Clinician-Administered PTSD Scale.

Table S1. Independent-samples *t*-tests were used to compare the two sub-groups in the EG (*n*=15 and *n*=5) and in the CG (*n*=8 and *n*=3). The EG participants who completed the study had significantly higher initial CAPS ratings (Comp vs Drop: *t*(18)=2.13, *p*<0.047). The mean initial CAPS ratings for the two EG sub-groups were 54.9 and 40.4, respectively (Table S1), yielding the effect size *d*=1.10. Among the CG participants who completed the study, the initial CAPS ratings were also higher than among those who dropped out, but not significantly (Comp vs Drop: *t*(9)=1.14, *p*<0.283). The mean initial CAPS ratings for the two CG sub-groups were 62.3 and 43.0, respectively (Table S1), with the effect size *d*=0.77. Also, the age difference between the two CG sub-groups trended toward significance (Comp vs Drop: *t*(9)=1.90, *p*<0.090). The other ratings in Table S1 did not show significant differences between the sub-groups. Therefore, the participants with higher initial PTSD severity (higher initial CAPS) were more likely to stay on and complete the study in either group (EG, CG).

*S2.2. LHIPS BOLD activity*

Mean fMRI percent signal changes for the LHIPS ROI (Fig. 2B) for the task runs across the three rtfMRI-nf sessions are exhibited in Fig. S2. They correspond to the LA BOLD activity results in Fig. 4. For the EG, no significant positive LHIPS BOLD activity was observed for any of the runs across the sessions (Fig. S2A). For the CG (Fig. S2B), the mean LHIPS BOLD activity trended toward significance after FDR correction for the Practice run in the 1st rtfMRI-nf session (PR: *t*(10)=2.81, *p*<0.018, *q*<0.092). When individual BOLD activity levels were averaged across the four nf runs (PR, R1, R2, R3), the group mean was not significant (*t*(10)=1.16, *p*<0.273) with effect size 0.35. Similarly, the mean LHIPS activity was positive and trended toward significance after the correction for Run 1 in the 3rd rtfMRI-nf session (R1: *t*(9)=2.46, *p*<0.036, *q*<0.061). The mean group activity for individual results averaged across the four nf runs in the 3rd session was not significant (*t*(9)=0.28, *p*<0.788) with effect size 0.09.

*S2.3. Amygdala FCS changes between sessions*

Fig. S3 examines correlations of the LA FCS changes between the final (3rd) and initial (1st) rtfMRI-nf sessions and the corresponding changes in PTSD severity (final vs initial CAPS ratings). The results are for the EG participants who completed the study (*n*=15). The whole-brain analysis is similar to the one for the LA fMRI connectivity changes between the sessions (Fig. 10), except that the FCS changes are considered instead. The cluster properties are specifies in Table S2. The correlation effects mapped in Fig. S3 are illustrated in Fig. S4. The FCS changes vs CAPS changes correlations are positive for the posterior nodes of the default mode network (DMN), including the PCC and the right angular gyrus (Figs. S3, S4, Table S2). This means that the stronger reduction in PTSD severity is associated with the stronger reduction in the FCS between the LA and the posterior DMN. Note that the left MiFG region (BA 6) in Figs. S3, S4, and Table S2 is *not* a part of the left DLPFC. This locus (−27, 5, 52) showed a negative, though non-significant, FCS vs initial CAPS correlation in the analysis reported in Fig. 8.

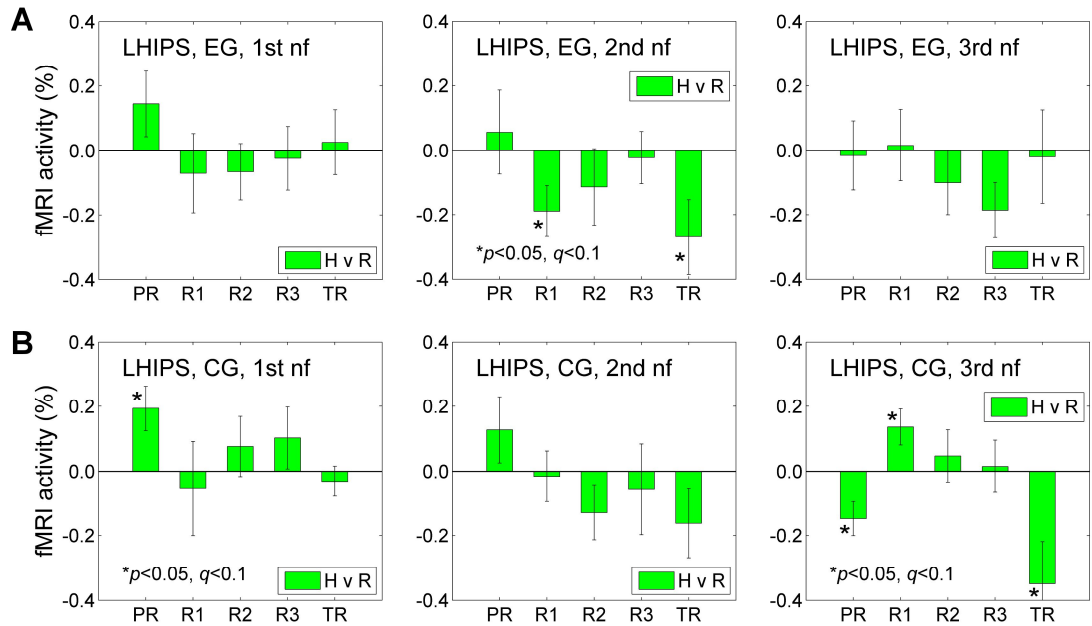

**Fig. S2.** BOLD activity of the left horizontal segment of the intraparietal sulcus (LHIPS) region during the Happy Memories conditions in the three rtfMRI-nf sessions. **A)** Average fMRI percent signal changes for the LHIPS ROI (Fig. 2B) across the five task runs in the 1st (visit 4), 2nd (visit 5), and 3rd (visit 6) rtfMRI-nf sessions (Fig. 1A) for the experimental group (EG). Each bar represents a mean GLM-based fMRI percent signal change for the Happy Memories conditions with respect to the Rest baseline (H vs R) in a given run, averaged across the group. The error bars are standard errors of the means (sem) for the group averages. The experimental runs and condition blocks are depicted schematically in Fig. 1B. **B)** Corresponding average fMRI percent signal changes for the control group (CG).

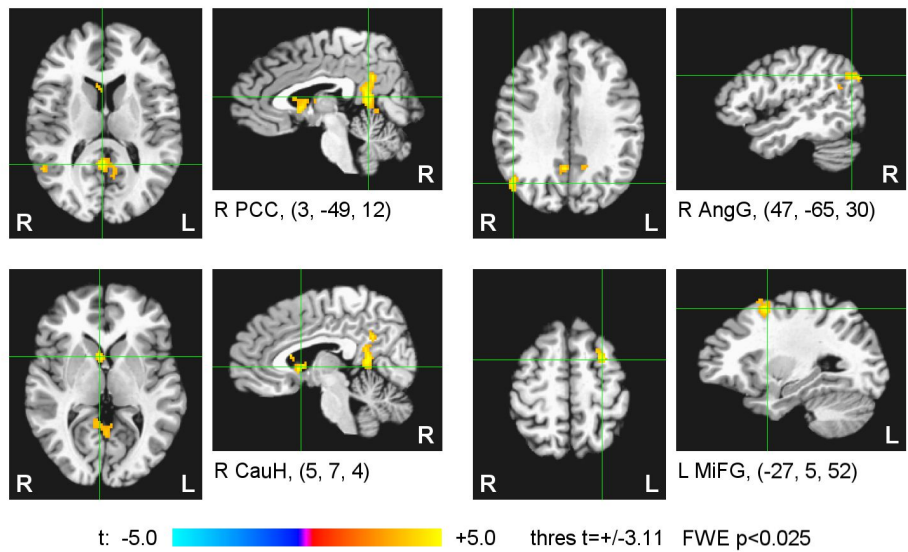

**Fig. S3.** Statistical maps of the correlation of the left amygdala fMRI connectivity slope (FCS) changes between the final (3rd) and initial (1st) rtfMRI-nf sessions and the corresponding changes in PTSD severity for the experimental group (EG). The correlation is a voxel-wise partial correlation with the changes in the CAPS ratings (final vs initial) controlled for corresponding changes in comorbid depression severity (HDRS) ratings and changes in LA FCS for central white matter. The maps are FWE corrected. The green crosshairs mark the statistical peak locations, with their Talairach coordinates specified underneath. Peak *t*-statistics values and the corresponding cluster properties are described in Table S2. Abbreviations: PCC – posterior cingulate cortex, AngG – angular gyrus, CauH – caudate head, MiFG – middle frontal gyrus.

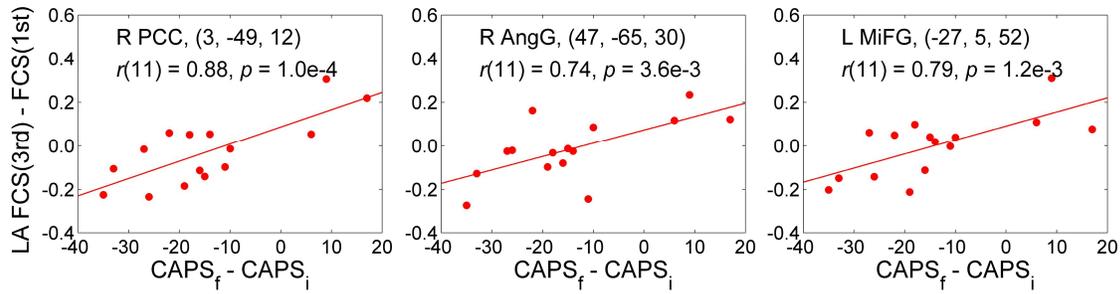

**Fig. S4.** Illustration of the correlation effects between the LA FCS changes and the corresponding PTSD severity changes for the EG, exhibited in Fig. S3. Each plot shows an average correlation effect for a 10-mm diameter ROI centered at a specified location. The correlation is a partial correlation with the changes in CAPS ratings (final vs initial) controlled for corresponding changes in HDRS ratings and changes in LA FCS for central white matter ($n$=15, $df$=11).

**Table S1.** Initial PTSD and depression severity ratings for the participants who completed the study and those who dropped out in the experimental and control groups. Abbreviations: CAPS – Clinician-Administered PTSD Scale; PCL-M – PTSD CheckList Military Version; HDRS – Hamilton Depression Rating Scale; MADRS – Montgomery-Asberg Depression Rating Scale.

|              | Experimental group (EG) | | Control group (CG) | |
| --- | --- | --- | --- | --- |
| **Measure**  | **Completed** mean (SD) | **Dropped out** mean (SD) | **Completed** mean (SD) | **Dropped out** mean (SD) |
| Participants | 15          | 5            | 8           | 3            |
| Age          | 30.8 (5.4)  | 31.4 (7.2)   | 36.8 (8.0)  | 27.0 (5.6)   |
| CAPS         | 54.9 (14.1) | 40.4 (8.9)   | 62.3 (22.4) | 43.0 (32.1)  |
| PCL-M        | 46.8 (10.9) | 38.2 (8.6)   | 49.5 (14.0) | 41.0 (28.6)  |
| HDRS         | 17.3 (6.7)  | 15.0 (2.3)   | 16.8 (9.1)  | 9.3 (7.6)    |
| MADRS        | 21.7 (9.8)  | 17.6 (4.6)   | 18.3 (14.1) | 14.0 (10.6)  |

**Table S2.** Correlation of the left amygdala fMRI connectivity slope (FCS) changes between the 3rd and 1st rtfMRI-nf sessions and the corresponding changes in PTSD severity (final vs initial CAPS). Location of the point with the peak group $t$-score and the number of voxels are specified for each cluster obtained after FWE correction for multiple comparisons.

| **Region** | **Late-rality** | **$x, y, z$ (mm)** | **$t$-score** | **Size (# voxels)** |
| --- | --- | --- | --- | --- |
| **FCS changes vs CAPS changes correlation** | | | | |
| Posterior cingulate cortex (BA 29) | R | 3, −49, 12 | 7.30 | 418 |
| Angular gyrus (BA 39) | R | 47, −65, 30 | 4.76 | 154 |
| Middle frontal gyrus (BA 6) | L | −27, 5, 52 | 5.60 | 140 |
| Caudate head | R | 5, 7, 4 | 5.78 | 84 |
| **Mean FCS changes** | | | | |
| not significant | | | | |

Notations: BA – Brodmann areas; L – left; R – right; $x, y, z$ – Talairach coordinates;
FWE corrected $p<0.025$ (Size – cluster size, minimum 81 voxels for uncorr. $p<0.01$).